\documentclass[structabstract]{aa}

\usepackage{amsmath,amssymb}
\usepackage{subfigure}
\usepackage{graphicx}
\usepackage[innercaption]{sidecap}
\usepackage{epstopdf}
\usepackage{color}
\usepackage{hyperref}
\usepackage{natbib} 
\bibpunct{(}{)}{;}{a}{}{,}

\newcommand{\Halo}{\mathcal{H}}

\newcommand{\ds}{\displaystyle}

\newcommand{\Mstar}{M_{\star}}
\newcommand{\Msun}{M_{\odot}}
\newcommand{\Zsun}{Z_{\odot}}
\newcommand{\sfg}{g^{\star}}

\newcommand{\EjTAB}{\tiny\textbf{Ej-TAB}}
\newcommand{\SMHTAB}{\tiny\textbf{SMH-TAB}}
\newcommand{\SNTAB}{\tiny\textbf{SN-TAB}}
\begin{document}

\title{Metal enrichment in a semi-analytical model, fundamental scaling relations, and the case of Milky Way galaxies} 

\authorrunning{M. COUSIN et al}

\titlerunning{Metal enrichment in SAM}

\author{M. Cousin \inst{1} \and V. Buat \inst{1} \and S. Boissier\inst{1} \and M. Bethermin \inst{2} \and Y. Roehlly\inst{1} \and M. G\'enois\inst{3}}

\institute{Aix Marseille Universit\'e, CNRS, LAM (Laboratoire d'Astrophysique de Marseille) UMR 7326, 13388, Marseille, France, e-mail : morgane.cousin@lam.fr \and European Southern Observatory, Karl-Schwarzschild-Str. 2, 85748 Garching, Germany \and Aix Marseille Universit\'e, Universit\'e de Toulon, CNRS, CPT, UMR 7332, 13288 Marseille, France}

\date{Received ? / Accepted ?}

\abstract{Gas flows play a fundamental role in galaxy formation and evolution, providing the fuel for the star formation process. These mechanisms leave an imprint in the amount of heavy elements that enrich the interstellar medium (ISM). Thus, the analysis of this metallicity signature provides additional constraint on the galaxy formation scenario.}{We aim to discriminate between four different galaxy formation models based on two accretion scenarios and two different star formation recipes. We address the impact of a bimodal accretion scenario and a strongly regulated star formation recipe on the metal enrichment process of galaxies.}{We present a new extension of the eGalICS model, which allows us to track the metal enrichment process in both stellar populations and in the gas phase. Based on stellar metallicity bins from 0 to $2.5\Zsun$, our new chemodynamical model is applicable for situations ranging from metal-free primordial accretion to very enriched interstellar gas contents. We use this new tool to predict the metallicity evolution of both the stellar populations and gas phase. We compare these predictions with recent observational measurements. We also address the evolution of the gas metallicity with the star formation rate (SFR). We then focus on a sub-sample of Milky Way-like galaxies. We compare both the cosmic stellar mass assembly and the metal enrichment process of such galaxies with observations and detailed chemical evolution models.}{Our models, based on a strong star formation regulation, allow us to reproduce well the stellar mass to gas-phase metallicity relation observed in the local universe. The shape of our average stellar mass to stellar metallicity relations is in good agreement with observations. However, we observe a systematic shift towards high masses. Our $M_{\star}-Z_g-$SFR relation is in good agreement with recent measurements: our best model predicts a clear dependence with the SFR. Both SFR and metal enrichment histories of our Milky Way-like galaxies are consistent with observational measurements and detailed chemical evolution models. We finally show that Milky Way progenitors start their evolution below the observed main sequence and progressively reach this observed relation at $z\simeq0$.}{}

\keywords{Galaxies: formation - Galaxies: evolution}

\maketitle

%*****************************
%
% Introduction
%
%*****************************

\section{Introduction}

For more than 30 years, cosmological models based on the $\Lambda$ cold dark matter(CDM) paradigm have explained the origin and evolution of structures in the Universe remarkably well \citep{Blumenthal_1984,Fu_2008}. However, the description of the baryonic processes acting at smaller scale (galaxies) remains problematic.

Gas flows and star formation are supposed to be the main actors of galaxy evolution \citep[e.g.][]{Dekel_2009a}. These mechanisms leave an imprint in the amount of heavy elements that are produced by stars and progressively enrich the ISM. A better understanding of both the gas phase and stellar metallicities is essential to better constrain the galaxy formation scenario. 

In the early 2000s, the analysis of hydrodynamic simulations allowed the formulation of a bimodal accretion scenario \citep[e.g.][]{Birnboim_2003, Keres_2005, Ocvirk_2008, Dekel_2009a, Dekel_2009b, Khochfar_2009, Brooks_2009, Faucher-Giguere_2011}. According to this bimodal scenario, the galaxies can grow through a cold and a hot mode. The smaller galaxies grow through a very efficient cold mode. At high redshift ($z > 2$), this mode is based on filamentary streams that feed the galaxy directly. In the hot accretion mode, the cosmological gas is shock heated in the halo, and then this hot gas has to cool before feeding the galaxy. The hot mode is predicted to complete and progressively replace the cold mode in more massive structures \citep{Faucher-Giguere_2011}. However, even when a hot isotropic atmosphere is present around a galaxy, some residual cold accretion can exist, in particular for massive halos at high redshifts \citep[e.g.][]{Thom_2011, Ribaudo_2011}. The recent works of \citealt{Nelson_2013}, based on a new generation of hydrodynamic codes, call into question this bimodal accretion scenario. According to this work, all the accreted gas could follow the hot mode. In this context, the existence of cold streams is questioned. 

In parallel to hydrodynamic simulations, semi-analytic models (SAMs) are  good tools for studying the impact of these different prescriptions. Since the pioneering work of \cite{White_1991}, the assumptions of i) a virial shock and ii) an isotropic hot atmosphere have been used in SAM \citep[e.g.][]{Cole_2000, Hatton_2003, Somerville_2008, Guo_2011, Henriques_2013}. A rapid and a slow cooling accretion mode can be identified in semi-analytical prescriptions, even if all the accreted gas is assumed to be shock heated \citep{Benson_2011}. These two cooling regimes are linked to the dynamical timescale of dark matter structures. In the case of low-mass structures the cooling rate is very efficient and the gas is rapidly accreted onto the galaxy. With this kind of model, the gas accretion cannot be followed in detail. In particular it is impossible to explicitly follow metals in the two accretion modes. Observations \citep[e.g.][]{Thom_2011, Ribaudo_2011, Crighton_2013} and simulations \citep[e.g.][]{Fumagalli_2011} indicate that the two accretion modes are associated with different metallicity contents. The cold mode corresponds to metal-poor gas, $log(Z/\Zsun) \le -2.0$. The hot gas is found with a higher metallicity, $log(Z/\Zsun) \ge -1.0$. The metals, observed in the hot gas surrounding galaxies are initially formed by stars in the galaxy. The enriched gas is then ejected by a set of feedback mechanisms in the form of supernovae (SN) and active galaxy nuclei (AGN).

These feedback processes are also essential to regulate the star formation activity in galaxies but they are still misunderstood \citep[e.g.][]{Bouche_2012}. Indeed, standard scenarios of regulation of  star formation lead to strong excesses of low-mass structures, mainly at high redshift ($z>3$). This standard scenarios based only on galactic fountains are gradually abandoned in favour of models using long re-accretion cycles \citep[e.g.][]{Guo_2011, Henriques_2013} or an  intermediate gas reservoir of non-star-forming gas that strongly regulates the star formation activity \cite{Cousin_2015a}.

In this work, we compare two accretion scenarios and two different star formation recipes. The first accretion scenario is described in detail in \cite{Cousin_2015b}. It assumes a bimodal accretion as initially proposed by \cite{Khochfar_2009} or \cite{Benson_2011}. This model is based on a cold and a hot reservoir. Both are fed by the metal-free cosmological accretion but only the hot reservoir receives  the galactic metal-rich ejecta in addition. As the metallicity of the wind phase depends on the galaxy metal enrichment process, the metallicity of the hot reservoir evolves with time. In the second accretion scenario, we assume a unique gas reservoir that receives both the metal-free cosmological accretion and  metal-rich gas ejected by the galactic feedback processes. In addition to these two accretion scenarios, we compare two different star formation recipes. First, we assume a standard star formation process where the gas is progressively converted into stars by assuming a given efficiency. Second, we use the recipe proposed by \cite{Cousin_2015a}: the freshly accreted gas is assumed to be non-star forming. It is progressively converted into star-forming gas and then into stars.

Our objectives are to characterise the impacts of these scenarios onto the galaxy metallicity signature in both stars and gas measured at $z\simeq 0$. Our work is based on a new chemodynamical package that we present here. Similar to its predecessors, \cite{Nagashima_2005}, \cite{Arrigoni_2010},  \cite{Yates_2013}, or more recently \cite{De_Lucia_2014}, our chemodynamical model is based on stellar evolution models. We assume a \cite{Chabrier_2003} IMF\footnote{We systematically use the \cite{Chabrier_2003} initial mass function (IMF), but the chemodynamical library is available for other IMF: \cite{Salpeter_1955}, \cite{Scalo_1998}, and \cite{Kroupa_2001}.}. The new library is built with a large set of initial stellar metallicties, including metal-free stellar populations. We also introduce a time-deleted SN feedback prescription, based on SNII and SNIa event rates. 

Our analysis is performed on the whole sample of our simulated galaxies and  on a sub-sample of star-forming (SF) galaxies. Observational measurements show that star formation rates and stellar masses are strongly correlated for SF galaxies \citep[e.g.][]{Noeske_2007, Elbaz_2007, Daddi_2007}. This tight correlation is called main sequence. Our sub-sample of SF galaxies is therefore defined in the context of this main sequence. 

The paper is organised as follows. In Sect. \ref{model} we describe the two accretion scenarios and give an overview of the other components of the model. In Sect. \ref{chemo_lib} we describe the new chemodynamical model used to follow the metal-enrichment process. In Sect. \ref{models_data_selections} we present the four different models and we describe the set of rules used to define our star-forming galaxy sample. In Sect. \ref{Galaxy_metal_enrichment} we present the metallicity signatures in both stars and gas phase measured at $z\simeq 0$ for our different scenarios. This first analysis allows us to define a best model among the four different prescriptions used. With this best model, we explore the link between the average metallicity of the gas and  star formation activity \citep[e.g.][]{Ellison_2008,Mannucci_2010,Lara-Lopez_2010}. In Sect. \ref{Miky_Way_like_galaxies} we focus on a sub-sample of galaxies: the Milky Way (MW) like galaxies. We analyse the impact of a set of selection criteria onto several properties such as the stellar and gas metallicity, the gas fraction, and the time elapsed since the last major merger. Then we detail their star formation and their metal enrichment histories. Finally, we replace the properties of the progenitors of MW-like galaxies in the context of the main sequence evolution. 
 The conclusion of this work is presented in Sect. \ref{conclusion}.  

\section{The eGalICS model}
\label{model}

We use the new eGalICS model described in detail in \cite{Cousin_2015b}. This semi-analytical model is applied on dark matter merger trees extracted from a pure N-body simulation. This simulation is based on a WMAP-3yr cosmology ($\Omega_m = 0.24$, $\Omega_{\Lambda} = 0.76$, $f_b = 0.16$, $h = 0.73$) and describes a volume of $[100/h]^3 Mpc$. In this volume, $1024^3$ particles evolve. Each particle has an elementary mass of $m_p = 8.536~10^7~\Msun$. We use the \verb?HaloMaker? code described in \cite{Tweed_2009} to identify haloes and their respective sub-structures (satellites) and build
the corresponding merger trees. In the merger tree evolution, we only consider halos containing, at least 20 dark matter particles. This limit leads to a minimal dark matter halo mass of $1.707\times10^9~\Msun$.

\begin{table}[h]
  \begin{center}
    \footnotesize{
      \begin{tabular}{lr}
        \hline
        Symbol & Definition and value \\
        \hline     
        $V_{wind}$               & wind velocity (due to SN) = $250$ km/s\\
        $\varepsilon_{\star}$    & star formation efficiency = 0.02\\ 
        $\varepsilon_{ej}$       & ejecta efficiency = 0.6\\ 
        $\varepsilon_{ff}$       & cold accretion efficiency = 1.0\\
        $\varepsilon_{cool}$     & cooling efficiency = 10.0\\
        \hline
      \end{tabular}}
  \end{center}  
  \caption{\footnotesize{Free parameters used in our models. We only list parameters whose values have changed in comparison to those given in \citealt{Cousin_2015a,Cousin_2015b}}}
  \label{free_parameters}
\end{table} 

\subsection{Dark matter and baryon accretion}

Dark matter haloes are fed following a smooth accretion mode. The dark matter accretion rate $\dot{M}_{dm}$ associated with a halo is built between two time-steps with the particles that are newly detected in the halo $\Halo$, and that have never been identified in an other halo before. The baryonic mass is then added progressively following this smooth, dark matter accretion,

\begin{equation}
\dot{M}_b = f_b^{ph-ion}(M_h,z)\dot{M}_{dm}
.\end{equation}

In this relation, $f_b^{ph-ion}(M_h,z)$ is the effective baryonic fraction that depends on the photoionisation model\footnote{We use the formulation proposed by \citealt{Gnedin_2000} and \cite{Kravtsov_2004}, but our effective filtering mass follows the expression given by \cite{Okamoto_2008}. We use a fit of the relation given in Fig. 15 of \cite{Okamoto_2008}. The slope index and redshift of re-ionisation are set to $alpha = 2$ and $z_{reion} = 7,$ respectively.}. 

From this global baryonic accretion ($\dot{M}_b$), we apply two different models. The first model is based on a single hot isotropic accretion. The second model assumes a bimodal accretion. The baryonic cosmological accretion is therefore divided into two phases. At a given redshift and for a given halo $\Halo$, the fraction of shock-heated gas ($f_{sh}(M_{vir},z)$) is computed following the \cite{Lu_2011a} prescription (their Eqs. 24 and 25)  
\begin{equation}
  \begin{split}
    f_{sh}(M_{vir},z) = \dfrac{1}{2}\left[0.1\times exp\left[-\left(\frac{z}{4}\right)^2\right]+ 0.9\right]~~\times \\
    ~~~\left[1+erf\left(\dfrac{logM_{vir}-11.4}{0.4}\right)\right].  
    \label{shock_heated_fraction}
  \end{split}
\end{equation}

We use two distinct reservoirs (cold and hot). The cold reservoir is fed by metal-free cosmological gas when the hot reservoir receives both metal-free cosmological gas and metal-enriched gas ejected by the galaxy. We assume that the gas ejected by the galaxy has no impact on the cold mode. While the cold reservoir feeds  the galaxy directly, the hot gas is assumed to reach the virial temperature and has to support the radiative cooling process before feeding the galaxy. 

The cooling rate is computed using the classical model proposed by \cite{White_1991}. The condensed mass enclosed in the cooling radius $r_{cool}$ is estimated assuming
\begin{itemize}
  \item{a hydrostatic equilibrium gas profile $\rho_g(r)$,}
  \item{a mean constant temperature $\overline{T}$, and}
  \item{a temperature and metal dependent cooling function $\Lambda(\overline{T},Z_g)$ \citep{Sutherland_1993}.}
\end{itemize}

The cooling rate of the hot atmosphere is then computed as
\begin{equation}
   \dot{M}_{cool} = \dfrac{\varepsilon_{cool}}{t_{dyn}}\int_0^{MIN(r_{cool},r_{vir})}\rho_g(r)r^2dr\, ,\end{equation}
where the dynamical time is adapted to the cooling radius\footnote{In the previous version, the dynamical time was computed using the virial radius. This new prescription, with a higher cooling rate efficiency, leads to a better transition between the cold-stream mode and cooling mode.} of the hot gas structure. The efficiency of the cooling process is governed by the free parameter $\varepsilon_{cool}$ (see Table \ref{free_parameters}).

\subsection{Star formation in disks}
\label{star_formation}

The accreted gas feeds the galaxy evolving in the centre of the dark matter halo. We assume that this cold gas initially forms a thin exponential disk. Gas acquires angular momentum during the mass transfer \citep{Peebles_1969}. After its formation, the disk is supported by this angular momentum \citep[e.g.][]{Blumenthal_1986, Mo_1998}.

We apply two different models for star formation. A standard model described in \cite{Cousin_2015b} and the recipes defined in \cite{Cousin_2015a}. In the standard prescription, the star formation is based on the empirical Schmidt-Kennicutt law \citep{Kennicutt_1998}. In this context we use a critical surface density threshold ($\Sigma_{crit} = 10\Msun pc^{-2}$, \citep{Schaye_2004}) and assume that only gas lying at surface density larger than this limit is available to form new stars\footnote{For a discussion on the impact of the surface density threshold parameter on the star formation process, please refer to \cite{Lu_2015}.}. In the second model, the accreted gas is assumed to be initially non-star forming. It is then progressively converted into star-forming gas and concentrates in the centre of the disk. We call $r_s$ the radius that encloses the star-forming gas. The SFR is given by
\begin{equation}
  (SFR)~~~\dot{M}_{\star} = \varepsilon_{\star}\dfrac{M_{\sfg}}{t_{dyn}}
  \label{star_formation_law}
,\end{equation}
where $\varepsilon_{\star}$ is a free efficiency parameter (see Table \ref{free_parameters}), $M_{\sfg}$ is the mass of star-forming gas. The star-forming timescale $t_{dyn}$ is then given by\begin{equation}
  t_{dyn} = \dfrac{r_s}{V_c}
  \label{dynamical_time}
,\end{equation}
where $V_c$ is the circular velocity of the disk computed at the radius $r_s$ by taking into the dark matter, the bulge, and the disk potential wells into account. 

\subsection{Supernovae feedback}
\label{sn_feedback}

We compute the ejected mass rate due to SN using kinetic energy conservation \citep{Dekel_1986}
\begin{equation}
        \dot{M}_{ej}V_{wind}^2 = 2\varepsilon_{ej}\eta_{sn}E_{sn}
        \label{sn_feedback_eq}
.\end{equation}
We use $E_{sn} = 10^{44} Joules$. The wind velocity is fixed and set to $V_{wind} = 250$ km/s, and $\varepsilon_{ej}$ is a free efficiency parameter. Unlike the previous version of the model, the ejecta rate is not directly linked to the SFR but to more realistic SNII and SNIa event rates $\eta_{sn}$, thus integrating all the stellar populations formed previously. The explicit computation of $\eta_{sn}$ is given in Sect. \ref{inst_sn_rate}. The realistic SNII and SNIa event rates $\eta_{sn}$, used in this new version, has modified the average mass-loading factor produced by SN feedback.  We increased the efficiency of the process to keep the same total SN feedback impact ($\varepsilon_{ej} = 0.6 > \varepsilon_{sn}f_{Kin,sn} = 0.3$).

\subsection{Mergers}

We define the merger type (major or minor) of two galaxies 1 and 2 using the following mass ratio, 
  \begin{equation}
    \eta_{merger} = \dfrac{MIN(M_{1/2,1}~;~M_{1/2,2})}{MAX(M_{1/2,1}~;~M_{1/2,2})}\,,
    \label{eta_merger}
  \end{equation}
where $M_{1/2}$ is the total mass (galaxy and dark matter halo) inside the galaxy half-mass radius. We consider the merger event as minor when $\eta_{merger} < 1/3$.

In the case of major merger ($\eta_{merger} > 1/3$), the galaxy structure and dynamics are completely modified. The stellar population of the remnant galaxy is a spheroid modelled by a \cite{Hernquist_1990} mass distribution. The gas components of the two progenitors are added and a new disk is formed. 

During a minor merger, we add  the stellar population of the bulges and disks separately to form the remnant bulge/disk. All cold gas reservoirs are added to the remnant disk.

To take  the strong modifications impacting galaxies during a merger into account, \cite{Cousin_2015a} defined a boost factor, $MAX\left[1,~\varepsilon_{boost}(\Delta t)\right]$, which increases the efficiencies of the star formation process. The boost factor is defined as 
\begin{equation}
    \varepsilon_{boost}(t) =
10^2\eta_{merger}\eta_{gas}exp\left(-\dfrac{t}{\tau_{merger}}\right).
    \label{boost_factor}
\end{equation}
This boost factor simulates a gas compression (decreasing with time) as a function of the gas content of the two progenitors, where $\eta_{gas}$ is the gas fraction in the post-merger structure, t is the time elapsed since the last merger event,  $\tau_{merger}$=0.05~Gyr is a characteristic merger timescale, and $\eta_{merger}$ is given in Eq.~\ref{eta_merger}. 

\section{The chemodynamical model}
\label{chemo_lib}

The impact of stars on the ISM-enrichment process is a function of their stellar mass ($m_{\star}$) and of their metallicity ($Z_{\star}$). We take  stellar masses into account between $m_{\star,min} = 0.1~\Msun$ and $m_{\star,max} = 100\Msun$. This scale has been chosen to match the stellar mass range of the BC03 spectrum library. 

This stellar mass range can be divided into three bins \citep{Romano_2005}: i) the very low-mass stars (VLMS) $[0.1-1[\Msun$, ii) the low- and intermediate-mass stars (LIMS) $[1-9]\Msun$, and iii) the massive stars (MS) $]9-100]\Msun$. For each mass bin, the metal production and the ISM re-injection processes are different. 

\begin{table}[h]
  \begin{center}
    \footnotesize{
      \begin{tabular}{l|c|c|c|c}
        \hline
        Metallicity & [1-3]$\Msun$ & [4-6]$\Msun$ & [7,9]$\Msun$ & [12-100]$\Msun$  \\
        \hline     
        $Z_{\star}^1 = 0.$      & C08 & GP13 & GP13 & H10\\      
        $Z_{\star}^2 = 0.0001$  & K10 & K10 & EXT & EXT \\
        $Z_{\star}^3 = 0.0004$  & INT & INT & P98 & P98 \\ 
        $Z_{\star}^4 = 0.004$   & K10 & K10 & P98 & P98 \\
        $Z_{\star}^5 = 0.008$   & K10 & K10 & P98 & P98 \\
        $Z_{\star}^6 = 0.02$    & K10 & K10 & P98 & P98 \\
        $Z_{\star}^7 = 0.05$    & EXT & EXT & P98 & P98 \\ 
        \hline
      \end{tabular}}
  \end{center}  
  \caption{\footnotesize{Reference of stellar evolution models used to defined ejected mass for the different initial stellar mass and initial stellar metallicities. K10: \cite{Karakas_2010}, P98: \cite{Portinari_1998}, C08: \cite{Campbell_2008}, H10: \cite{Heger_2010}, and GP13: \cite{Gil_Pons_2013}. Interpolated data and extrapolated data are mentioned with INT and EXT, respectively.}}
  \label{chemo_model_ref}
\end{table} 

\begin{figure*}[t]
  \includegraphics[scale=0.52]{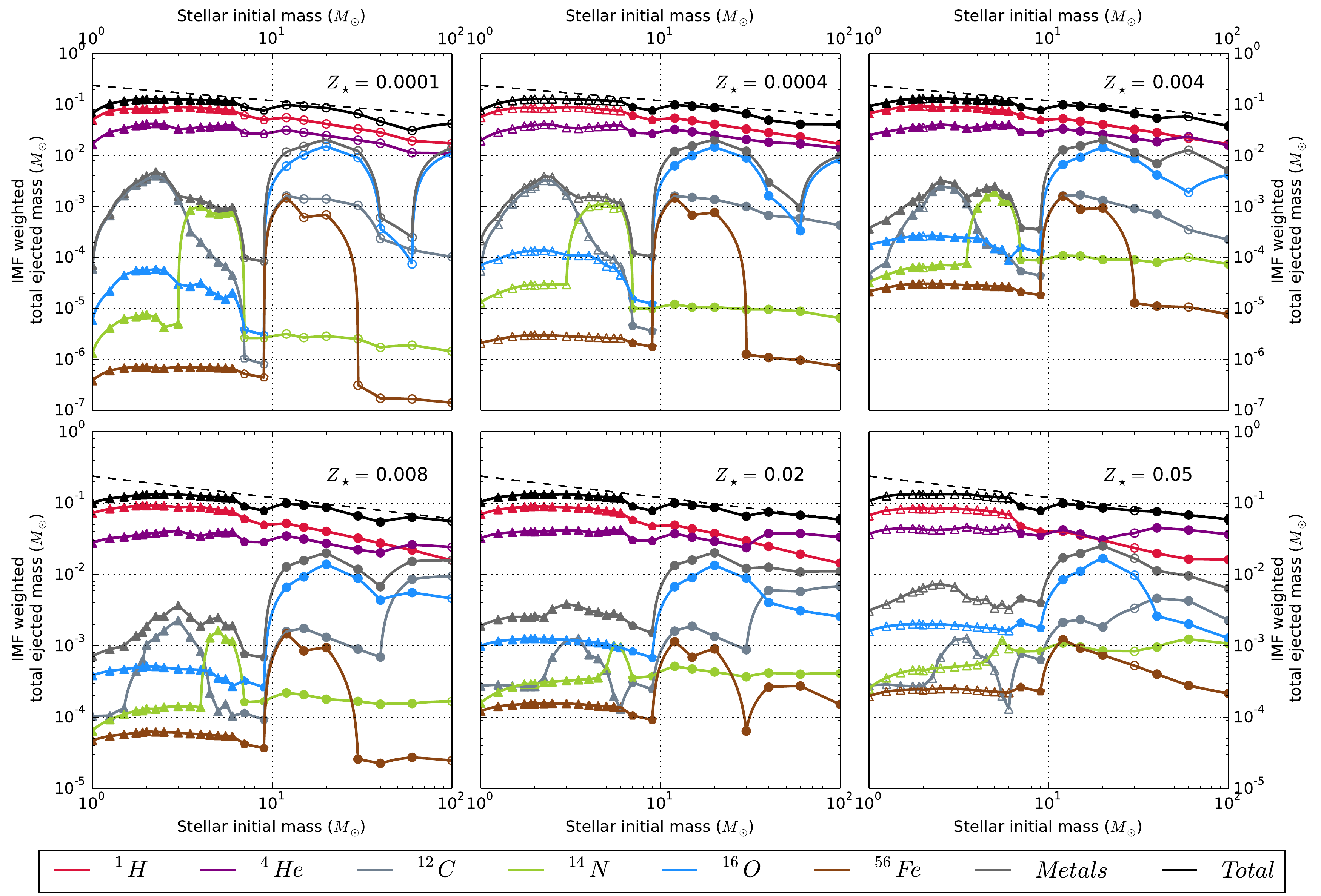}
  \caption{\tiny{IMF-weighted ejected mass for stars with initial mass in the range [1.0:100] $\Msun$. The ejected mass is given for six different stellar metallicities: $Z_{\star} = [0.0001,~0.0004,~0.004,~0.008,~0.02,~0.05];$ and the ejected mass is provided for six main ISM elements: $^1H$ (red), $^4He$ (purple), $^{12}C$ (light grey), $^{14}N$ (green), $^{16}O$ (blue), and $^{56}Fe$ (brown). The total ejected mass (black) and the metal ejected masses (dark grey) are also given. The black dashed line shows the IMF-weighted initial mass of the star. Ejected masses for LIMS ($m_{\star} \in [1,~9]\Msun$), follwoing \cite{Karakas_2010} and \cite{Portinari_1998}, are plotted with triangles and pentagons, respectively. Ejected masses for MS ($m_{\star} > 9\Msun$) are plotted with circles. Filled symbols show original data, while open symbols are dedicated to interpolated (extrapolated) data.}}
  \label{tot_Ejecta_plot}
\end{figure*}

The new chemodynamical library is built by following six of the main ISM elements $^1H$, $^4He$, $^{12}C$, $^{14}N$, $^{16}O,$ and $^{56}Fe$. The production and  re-injection processes are associated with stars with initial mass between $m_{\star,min}$ and $m_{\star,max}$, and for seven metallicity\footnote{The metallicity $Z$ is defined here as the metal mass fraction. In this work the solar metallicity is set to $\Zsun = 0.02$.} bins listed in Table \ref{chemo_model_ref}. These bins are able to cover gas composition from metal-free to super-solar metal fraction. We built this metallicity scale by merging different sets of stellar evolution models: \cite{Portinari_1998}, \cite{Campbell_2008}, \cite{Karakas_2010} and \cite{Heger_2010}, and \cite{Gil_Pons_2013} (see Table \ref{chemo_model_ref}). Some more recent library of massive stars yields are currently available (e.g. Geneva group). These works have shown that C,N, O, for example are affected by mass loss and rotation. The current work is dedicated to the chemical evolution of galaxies, but we have also included in parallel the prediction of stellar population spectra via the BC03 spectra library. To stay in agreement with the BC03 library, we based our yield library on the Padova model and we have therefore used the same metallicity scale.

\subsection{Very low-mass stars (VLMS)}

The VLMS contains stars with mass between $m_{\star} = 0.1~\Msun$ and $m_{\star} = 1.0\Msun$. We assume that VLMS are eternal. They simply remove their mass from the gas evolution cycle. 

\subsection{Low- and intermediate-mass stars (LIMS)}

The LIMS are stars with a initial mass between $m_{\star} = 1\Msun$ and $m_{\star} = 9\Msun$. They end their life as white dwarfs (WD) after passing through a phase of double-shell burning, known as the thermally pulsing asymptotic giant branch (TP-AGB) phase. During this phase He, C, and N are convected to the outermost layers and re-injected into the ISM. We assume that all the mass ejected by a LIMS is injected in the ISM at the end of the stellar life\footnote{The stellar lifetime is a function of both the stellar mass and metallicity. We use the metallicity-dependent lifetimes of \cite{Portinari_1998}. As suggested by, for example \cite{Padovani_1993}, for a star with a mass smaller than $m = 0.5$, and for all metallicity bin, we set $\tau_{m<0.5} = \tau_{m=0.5}$.} of a star of initial mass $m_{\star}$. Ejecta processes coming from LIMS are taken into account following \cite{Karakas_2007} and \cite{Karakas_2010}. All their models were evolved from the zero-age main sequence to the tip of the TP-AGB. The details of this procedure is described in \cite{Karakas_2009} and references therein. In these works, ejected masses are given for a range of initial stellar masses, $m_{\star} = [1-6]~\Msun$, and a set of metallicities, $Z_{\star} = [0.0001,~0.004,~0.008,~0.02]$. These metallicity bins correspond to the $Z_{\star}^2$, $Z_{\star}^4$, $Z_{\star}^5$, and $Z_{\star}^6$ bins of our chemodynamical library. To complete the metallicity set, we interpolated(extrapolated) the total ejected mass produced by LIMS on a metallicity log scale for metallicity bin $Z_{\star}^3$($Z_{\star}^5$). To complete the initial stellar mass range given by \cite{Karakas_2010}, $[1-6]~\Msun$, we take ejected mass coming from stars with intermediate masses $m_{\star} = 7\Msun$ and  $m_{\star} = 9\Msun$  into account following \cite{Portinari_1998}. 

\subsection{Massive stars (MS)}

The MS correspond to $m_{\star} > 9\Msun$. They represent only a small fraction of the stellar mass, i.e. $21\%$ for a \cite{Chabrier_2003} IMF \citep[e.g.][]{Romano_2005, Romano_2010}. Nevertheless, MS are very important contributors to the ISM enrichment process. Their nucleosynthesis produces heavy elements that are directly re-injected into the ISM through stellar winds during  their entire life. MS end their life in a type II supernovae episode (SNII) during which a large amount of energy and the majority of the newly synthesised metals are injected into the ISM. Ejecta from MS are taken into account following \cite{Portinari_1998}. This choice is based on the large available initial stellar mass range they provide, i.e. $m_{\star} = [12,~120]~\Msun$. \cite{Portinari_1998} also take  the mass loss into account through winds prior to the SNII episode. Stellar ejecta have been computed for the following metalicities: $Z_{\star} = [0.0004,~0.004,~0.008,~0.02,~0.05]$. To complete our metallicity scale, ejecta associated with MS with $Z = 0.0001$ were extrapolated on a metallicity log scale. As for LIMS we consider that the elements released during the SNII phase are instantaneously injected into the ISM at a time $t$ corresponding to the lifetime of the star $t = \tau_m$. Before the SNII episode, we also assume a continuous wind ejecta process acting during a time lapse equal to the lifetime of the star $\Delta t = \tau_m$. 

\subsection{Total ejected masses, the overall mass range}
\label{stars_ejecta}

\begin{figure*}[ht!]
  \includegraphics[scale =0.52]{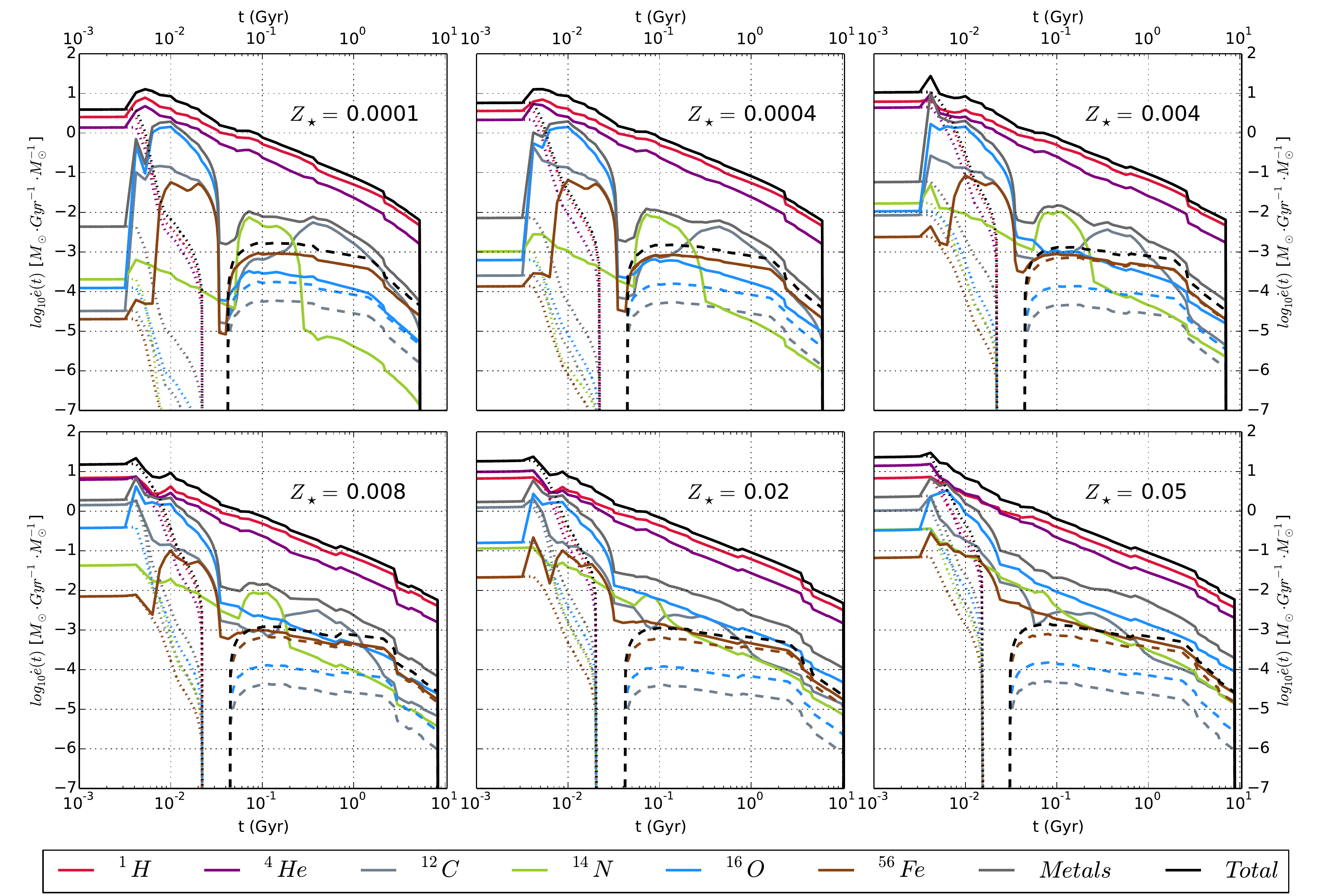}
  \caption{\tiny{Instantaneous ejecta rates computed for a SSP of $1\Msun$ and a Chabrier IMF. The contribution of the different main ISM elements are plotted with coloured solid lines: $^1H$ (red), $^4He$ (purple), $^{12}C$ (light grey), $^{14}N$ (green), $^{16}O$ (blue), and $^{56}Fe$ (brown). The total ejected mass and  metal ejected mass are shown in black and dark grey, respectively. The contribution of wind ejecta and SNIa are plotted with dotted lines and dashed lines, respectively.}}
  \label{Ejecta_rates_plot}
\end{figure*}

Figure \ref{tot_Ejecta_plot} shows the evolution of the IMF-weighted total ejected mass for six main ISM elements ($^1H$, $^4He$, $^{12}C$, $^{14}N$, $^{16}O,$ and $^{56}Fe$). The metal ejected mass and total ejected mass are also plotted. 

In the LIMS range, $[1,~9]\Msun$, a good complementarity is found between ejected masses coming from \cite{Karakas_2010} $[1-6]~\Msun$ and \cite{Portinari_1998} $[7,9]~\Msun$.

In the different panels, dedicated to different metallicities, we note a gradual increase of the metal ejected mass with the metallicity. For example, solar metallicity ($Z_{\star} = 0.02$) LIMS eject, on average, ten times more oxygen than LIMS with metallicity $Z_{\star} = 0.0004$. 

\subsection{Event rates}
\label{event_rates}

To convert the mass of gas ejected by each star (according to its initial mass) in a continuous ejection rate over time, we have to build the event rates associated with each specific ejection process.

\subsubsection{TP-AGB and Type II supernovae events}

For LIMS we assume that the mass is ejected during the TP-AGB phase. For MS the majority of the mass is ejected during the type II supernovae episode. The remaining part is continuously ejected by the stellar winds during the whole life of the MS. The TP-AGB and the Type II SN event rates are therefore given by the death rate of their associated stars. This death rate is given by the derivative of the inverse function of the stellar lifetime $\tau_{m}$, \begin{equation}
\ds \eta_{\star}(t) = \dfrac{\varphi(m)}{m}\left|\dfrac{dm}{d\tau_{m}}\right|
\label{eta_AGB_SNII}
,\end{equation}
where $\varphi(m)$ is the IMF. For stars with masses larger than $1\Msun$ the stellar lifetime is a decreasing function of the stellar mass. With this formulation, the injection process associated with a single stellar population (SSP) begins with the contribution of the most massive stars ($m_{\star} = 100\Msun$). It is then progressively fed by less and less massive stars ($m_{\star} \rightarrow 1\Msun$).   

\subsubsection{Type Ia supernovae contribution}

In parallel to the TP-AGB phases of LIMS and SNII explosion of MS, the ISM is also enriched by type Ia supernovae explosions coming from binary systems. The observed properties of SN Ia suggest that these systems may be produced by the thermonuclear explosion of accreting white dwarfs (WD), even if the exact binary evolution has not yet been identified \citep[e.g.][]{Greggio_1983, Matteucci_1986, Nomoto_1995, Wheeler_1995, Matteucci_2001}. 

We compute the SNIa contribution according to the model proposed by \cite{Greggio_1983}. The mass ejected for each of the six elements that we followed are given by the model of \cite{Iwamoto_1999} (W7).  

The SNIa event rate following a burst of star formation and leading to a stellar mass  of $1\Msun$ (SSP) is given by
\begin{equation}
\ds \eta_{snIa}(t,Z_{\star}) = \mathcal{A}\left|\dfrac{dM_2}{d\tau_m}\right|\int_{M_{B,inf}}^{M_{B,sup}}f(\mu)\varphi(\mu)\dfrac{dM_{B}}{M_B^2}
\label{eta_SNIa}
,\end{equation}
where the integration limits $M_{B,inf}$ and $M_{B,sup}$ are given by
\begin{equation}
\ds M_{B,inf} = MAX\left(2M_2(t),M_{Bm}\right)
\end{equation}
and 
\begin{equation}
\ds M_{B,sup} =\frac{1}{2}M_{BM}+M_2(t)
.\end{equation}
In these previous expressions, $M_B = M_1 + M_2$ is the total mass of the binary system. We assume that this mass is enclosed between $M_{BM}=16\Msun$ and $M_{Bm}=3\Msun$, and $M_2$ is the mass of the companion. At a given time $t$ after the formation, $M_2$ is the mass of the LIMS that died at $t$ and is therefore given by the inverse function of the stellar lifetime $m(\tau_m=t)$. As the stellar lifetime is a function of the stellar metallicity, the SNIa event rate is impacted by the stellar metallicity. In Eq. \ref{eta_SNIa}, $\mu=M_2(t)/M_B$ is the companion mass ratio and $f(\mu) = 24\mu^2$ is the distribution function of binary systems. Finally, $\mathcal{A} = 0.03$ is the fraction of binary systems with the right characteristics to become SNIa. 

\subsection{Ejecta rates for a SSP}

We introduce here the expression of the time dependent ejecta rate, $\dot{e}(t)$, generated by a SSP. This formulation is the elementary step used in the computation of the evolution of a more complex stellar population.

The time dependent ejecta rate is given by the sum of three different terms. Each of these terms corresponds to a specific enrichment process, i.e. 
\begin{equation}
  \begin{split}
    \dot{e}(t,Z_{\star},elt) & = m_{ej}(m_{\star},Z_{\star},elt)\eta_{\star}(t) \\
    & + m_{ej}(snIa,elt)\eta_{snIa}(t) \\
    & + \varphi(m_{\star})\int_{9\Msun}^{m_{\star}}\frac{m_{ej,w}(m',Z,elt)}{\tau(m')}dm'
    \label{ejecta_rate_eq}
  \end{split}
.\end{equation}

In this expression, the first term corresponds to the mass of element $elt$ ejected by stars of initial mass $m_{\star}$ and initial metallicity $Z_{\star}$ at the end of their life. Stars of smaller and lower mass $m_{\star}$ die progressively, each ejecting gas into the ISM. $m_{ej}(m_{\star},Z_{\star},elt)$ represents the contribution of LIMS during their TP-AGB phase or the contribution of the MS during the SNII explosion. The second term represents the contribution of SNIa. It is based on $m_{ej}(snIa,elt)$, which is the mass of the element $elt$ ejected by a type Ia supernovae. The third term of Eq. \ref{ejecta_rate_eq} represents the MS wind contribution. We assume that the mass ejected through these winds are continuously injected in the ISM during the stellar lifetime, $\Delta t = \tau_m$. $m_{ej,w}(m_{\star},Z_{\star},elt)$ is the mass of the element $elt$ ejected through stellar winds by a MS of initial mass $m_{\star}$ and initial metallicity $Z_{\star}$. $\eta_{\star}$ is the instantaneous stellar death rate and $\eta_{snIa}$ the instantaneous SN rate.

Fig. \ref{Ejecta_rates_plot} shows the evolution of the ejecta rates associated with a SSP of $1\Msun$ for different metallicity bins. These rates are given for the main ISM elements. The initial ejecta rate corresponds to the contribution of winds from massive stars. This plateau ends at the death of the most massive star ($m_{\star} = 100\Msun $).

\subsection{Metal-free primordial interstellar medium}

\begin{figure}[t]
  \begin{center}
    \flushleft \includegraphics[scale =0.45]{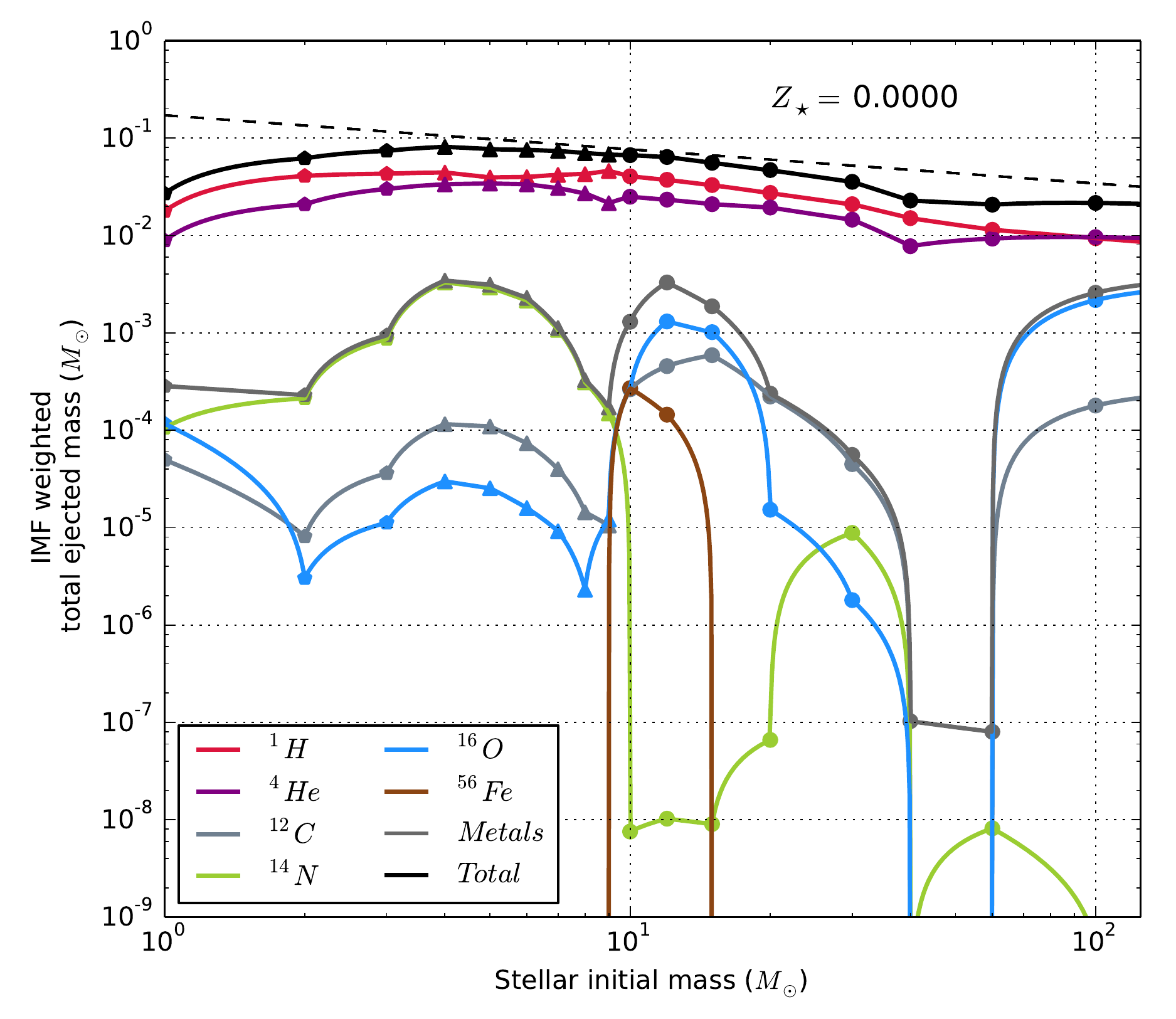}
    \center \includegraphics[scale =0.44]{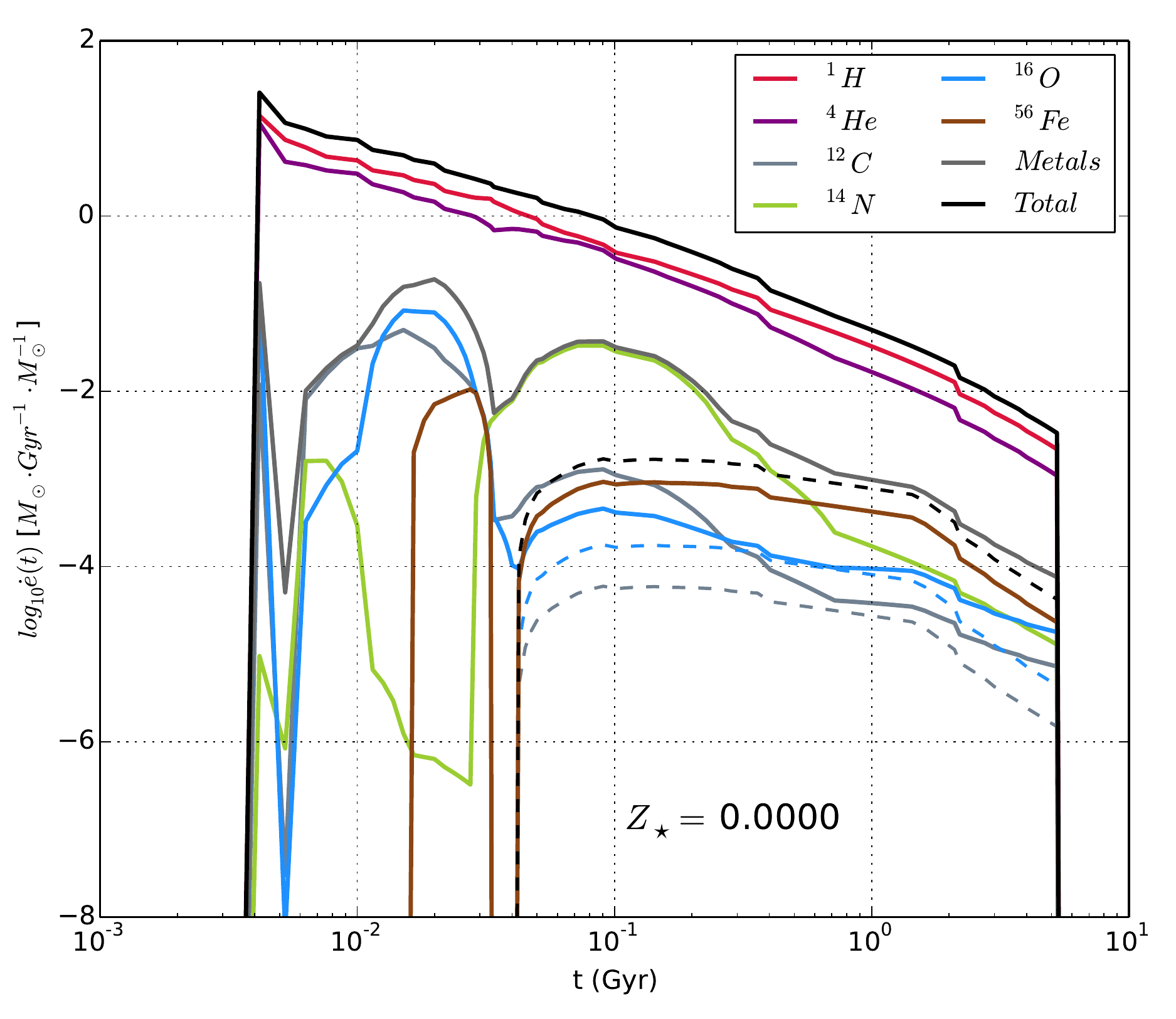}
  \caption{\tiny{\textbf{Upper panel}: The IMF weighted ejected mass for metal-free stars with initial masses in the range [1:100]$\Msun$. Ejected mass are given for six main ISM-elements  as in Fig. \ref{tot_Ejecta_plot}. The total ejected mass (black) and the metal ejected mass (dark grey) are also given. \cite{Campbell_2008} model is plotted with pentagons, \cite{Gil_Pons_2013} model with triangles and \cite{Heger_2010} model with circles. \textbf{Lower panel}: Time evolution of the ejecta rate for a metal-free SSP of 1$\Msun$. The specific contribution of the SNIa process to the ISM enrichment is plotted with dashed-line.}}
  \label{Ejecta_Z0_plot}
  \end{center}
\end{figure}

During the first steps of a galaxy formation process, the gas accreted onto the disk is metal-free. To take this feature into account, we have to build a metal-free ($Z_{\star}^1 = 0.$) bin in our chemodynamical model. We use stellar evolution models from \cite{Campbell_2008} and \cite{Gil_Pons_2013} for LIMS and from \cite{Heger_2010} for MS. The instantaneous ejecta rate is given by Eq. \ref{ejecta_rate_eq} but without the MS wind term; the \cite{Heger_2010} model does not take an explicit wind contribution into account. For metal-free MS, all the ejected mass is injected into the ISM during the SNII phase.

The upper and lower panels of Fig. \ref{Ejecta_Z0_plot} show the IMF-weighted total ejected mass and the time evolution of the instantaneous ejecta rates computed for a metal-free SSP of $1\Msun$. Without the MS wind contribution, there is no plateau at the beginning of the time evolution of the ejecta rates. Gas and metals are injected only after the death of the most massive stars. For SNIa, the ejected mass is still deduced from the W7 model of \cite{Iwamoto_1999}.

\subsection{From single stellar populations to realistic star formation histories}

During its evolution a galaxy forms stars in a continuous process. All these populations have to be followed in all the galaxies. The formation and evolution of the stellar population are taken into account  using a set of history tables described in Appendix \ref{Histtables}.

\section{Models and sample selection criteria}
\label{models_data_selections}

\subsection{The models}

We generated four different models, noted $m_1$ to $m_4$. We based $m_1$ and $m_3$  on the isotropic hot accretion paradigm,  and $m_2$ and $m_4$  on bimodal accretion in which small structures are essentially fed by cold metal-free gas. We based $m_1$ and $m_2$  on the standard star formation process \citep{Cousin_2015b} while, in $m_3$ and $m_4$, the new stars are formed following the recipe of \cite{Cousin_2015a} (see \ref{star_formation}). The different configurations are listed in the table \ref{model_definitions}.

\begin{table}[h]
  \begin{center}
    \footnotesize{
      \begin{tabular}{lcr}
        \hline
        Model & Accretion & star formation \\
        \hline     
        $m_1$  & isotropic (hot only) & standard \\
                $m_2$  & bimodal (hot+cold)   & standard \\
                $m_3$  & isotropic (hot only) & non-star-forming gas (non-sfg)\\
                $m_4$  & bimodal (hot+cold)   & non-star-forming gas (non-sfg) \\
        \hline
      \end{tabular}}
  \end{center}  
  \caption{\footnotesize{Definition of our four different models. As explained in the text, these four models are based on two different scenarios of gas accretion and star formation processes.}}
  \label{model_definitions}
\end{table} 

The values of the star formation efficiency, $\varepsilon_{\star}$, and the ejecta efficiency, $\varepsilon_{sn}$, are the same in all models (Table \ref{free_parameters}).

\subsection{Sample selection: Star-forming galaxy sample}

Star-forming galaxies (SF) are the subject of many observational studies. These objects are key elements of the stellar mass assembly. The relation found between their stellar mass and their SFR (main sequence) argues for a secular evolution. This relation is the main topic of a large number of studies \citep{Noeske_2007, Elbaz_2007, Daddi_2007}.

\begin{figure}[th]
  \includegraphics[scale=0.6]{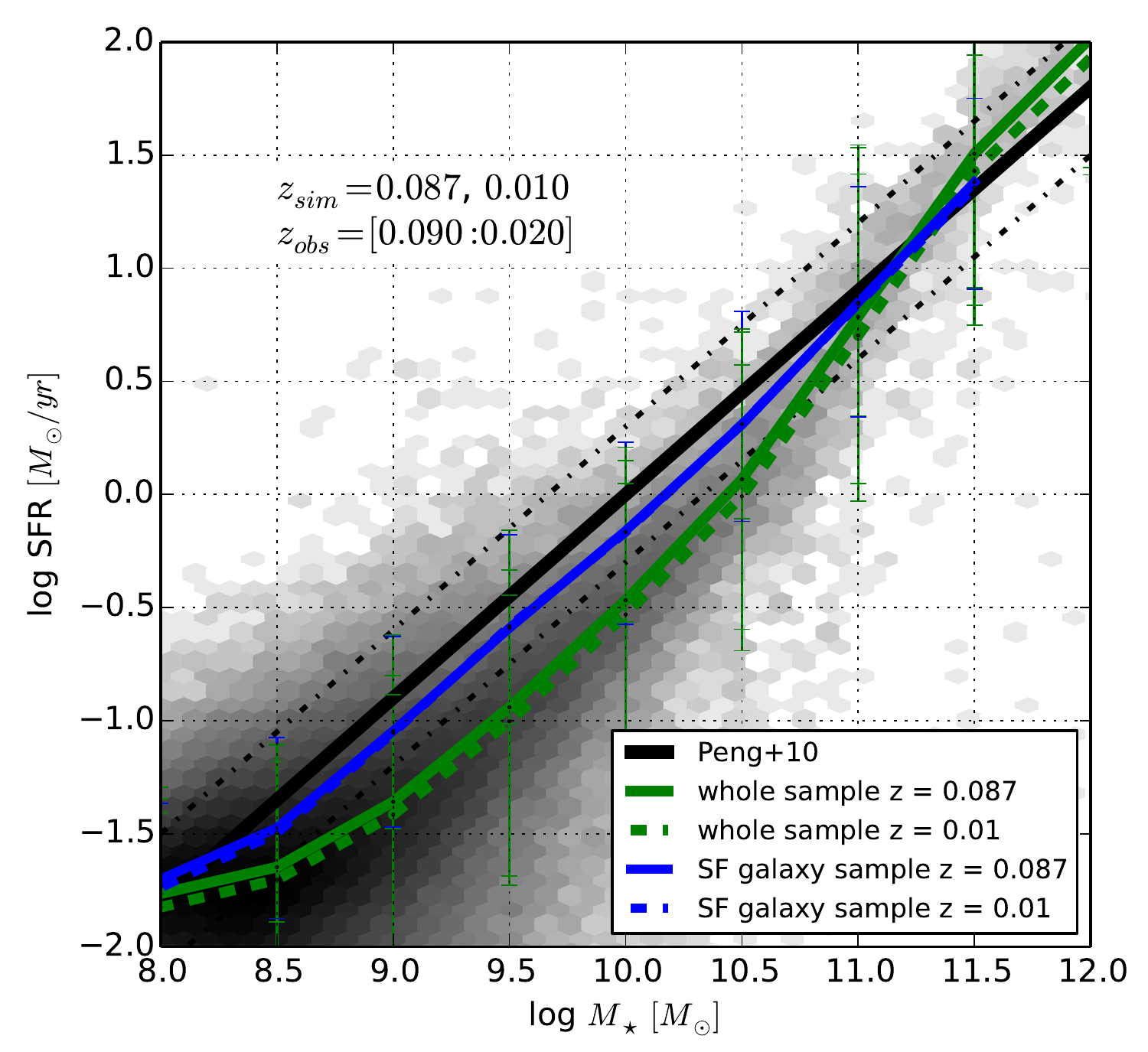}
  \caption{\tiny{ $\Mstar$-SFR relation at $z=0.01$ and $z\simeq 0.09$. The grey scale histogram materialises the population of the whole galaxy population extracted at $z\simeq 0.09$ from our model $m_3$. The mean trends of our whole and sub-sample of SF galaxies are shown with green and blue lines, respectively. Solid and dashed lines show the trend at $z=0.01$ and $z\simeq 0.09,$ respectively. Error bars are computed following the first and the third quartile of the distribution. These relations are compared to the observational data of \cite{Peng_2010}: the black solid line shows the mean trend of the main sequence found in the SDSS-zCOSMOS ($0.02 < z_{obs} < 0.09$) survey. The two dashed lines materialise a scatter of 0.3 around the mean relation.}}
  \label{Main_sequence_z0}
\end{figure}

The tight correlation between the stellar mass and the SFR, leads to the measure of a specific SFR (sSFR = $SFR/\Mstar$) that slowly evolves with the stellar mass. We define our sub-sample of SF galaxies using a sSFR range that evolves with the stellar mass. We use the average relation measured by \cite{Peng_2010} in the SDSS-zCOSMOS survey (their Fig. 1), i.e.
\begin{equation}
        log~sSFR = -10\pm 0.3 - 0.1(log~M_{\star} -10) 
.\end{equation}
From this average relation, at a given stellar mass, the sSFR range is defined by applying a scatter of 0.3 dex corresponding to the standard dispersion measured around the main sequence by, for example \cite{Salim_2007}, \cite{Elbaz_2007}, or \cite{Peng_2010}. 

We add a set of rules to the sSFR criterion to define our sub-sample of SF galaxies. To be included in the sample, a galaxy must:
\begin{itemize}
        \item{be hosted by a dark matter halo with $M_{vir}>10^{10}\Msun$ and}
        \item{not have suffered a major merger during the last 0.05 Gyr.}
\end{itemize}

The first criterion reduces the effects of resolution in mass. We recall that the minimum halo mass is $M_h^{min} = 1.707\times10^9~\Msun$. 

During a (major-)merger, a SF-galaxies leaves the main sequence for a short time. During this period the SFR of a merger galaxy is larger than that of a main sequence galaxy of similar stellar mass. The last criterion allows us to remove those galaxies that have recently suffered a major merger.  

Fig. \ref{Main_sequence_z0} shows the $\Mstar$-SFR relation extracted at $z=0.01$ and $z\simeq 0.09$ from our model $m_3$ for the whole sample and for our SF galaxy sample. These results are compared to observational data of \cite{Peng_2010} based on the SDSS-zCOSMOS survey ($0.02 < z_{obs} < 0.09$). As illustrate in Fig. \ref{Main_sequence_z0} the predictions of our model extracted at the two limits of the \cite{Peng_2010} sample are very close and almost indistinguishable.

Comparing to the observed relation, we note that our SFRs are lower in low-mass objects ($\Mstar < 10^{10}\Msun$), whereas the star formation activity is higher in massive objects ($\Mstar > 10^{11}\Msun$). The gap is strongly reduced for our sub-sample of SF galaxies. Concerning the high-mass range ($\Mstar > 10^{11}\Msun$), it is possible to invoke the activity of a super-massive black hole (SMBH) to reduce gas accretion onto the galaxy and reduce  star formation activity \citep[e.g.][]{Croton_2006,Bower_2006,Somerville_2008}. As explained in \cite{Cousin_2015b}, the impact onto the cooling process of the SMBHs in our model is weaker than in other models and does not sufficiently reduce star formation activity in massive galaxies. In the low-mass regime ($\Mstar < 10^{10}\Msun$), the galaxies have star formation activity that is too low. The difference between the mean trend of our whole sample and the observed trend is close to 0.4 dex. These low SFRs are due to a very efficient regulation of the star formation process (in this case, the non-sfg reservoir). This regulation must be strong  to reproduce the faint end of the stellar mass function. We point out here the difficulty of simultaneously matching   the stellar mass contents of the galaxies and the rhythm of this stellar mass assembly. This discrepancy is also highlighted by \cite{Henriques_2015}. Indeed, even if the last version of the Munich model leads to an excellent agreement on the stellar mass function, the cosmic SFR density is lower than the observed data points. Their main sequences also seem to show a SFR that is too low for the lowest galaxies.

\section{Galaxy metal enrichment}
\label{Galaxy_metal_enrichment}

In this section, we compare the behaviour of metallicity in our model to local ($z\simeq 0.09$) data on the stellar mass to stellar metallicity relation \ref{stars_Z}, on the stellar mass to gas-phase metallicity \ref{gas_Z}, and on the relation with the SFR (\ref{SFR-Zg-Ms})

\subsection{Stellar metallicity}
\label{stars_Z}

\begin{figure*}[t]
  \begin{center}
    \includegraphics[scale = 0.65]{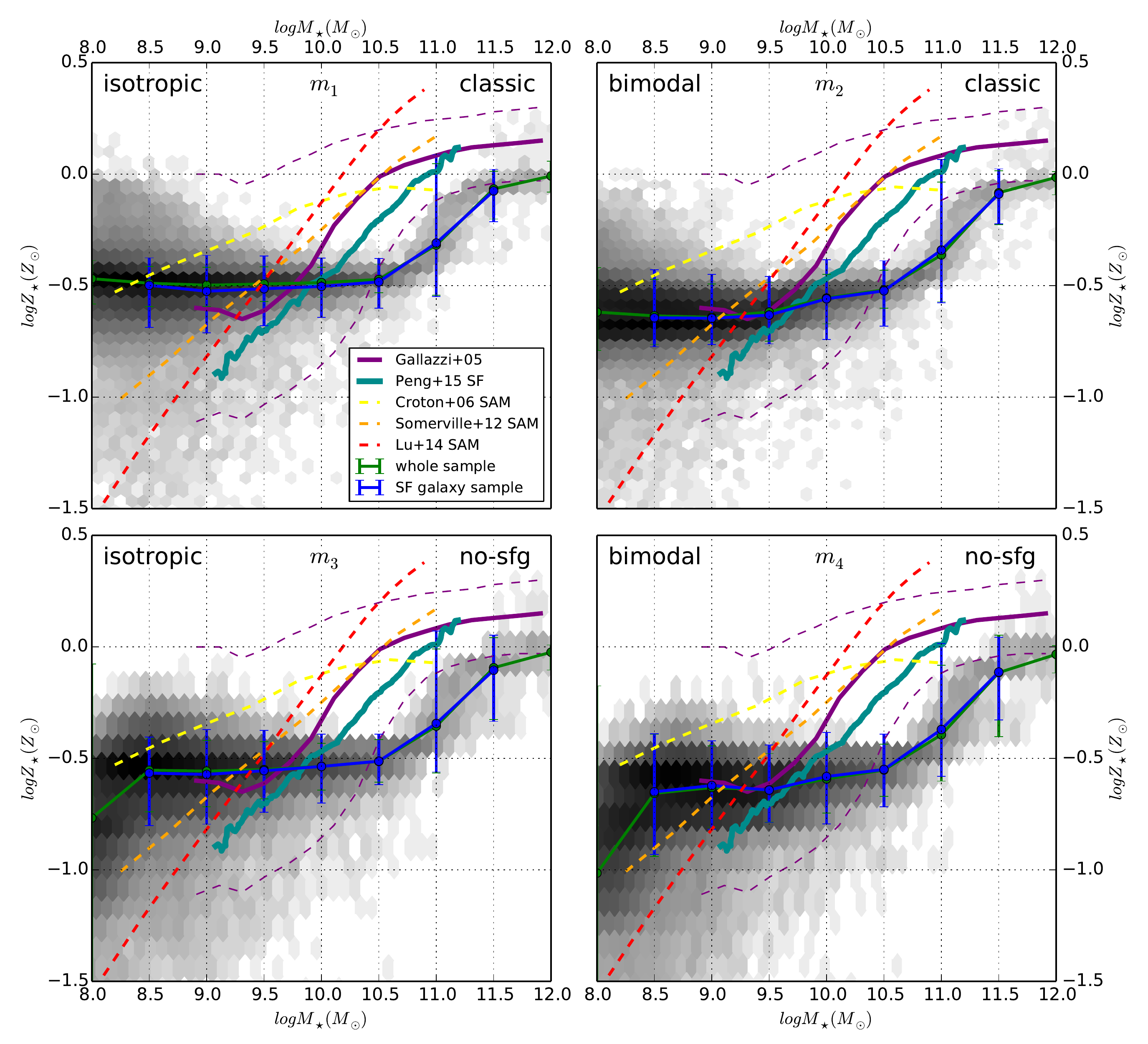}
        \caption{\tiny{$M_{\star}-Z_{\star}$ relation. The stellar metallicity in our models is computed using a bolometric luminosity weighting. Four panels correspond to the four different scenarios (model $m_1$ to $m_4$ see Table \ref{model_definitions} for model references). In all panels, the grey scales are built with all the galaxies at $z = 0.09$. The green and blue lines show the median trend of our whole sample and our SF galaxy sample, respectively. Error bars are computed with the first and the third quartile of the distribution. Our models are compared to observations from \cite{Gallazzi_2005} (grey lines). We also compare our results to a set of other SAM: \cite{Croton_2006} (yellow dashed line) \citealt{Somerville_2008,Somerville_2012} (orange dashed line), and \cite{Lu_2011b} (red dashed line). The average relations for the different SAMs were extracted from \cite{Lu_2014}.}}
  \label{Ms_Zs}
  \end{center}
\end{figure*}

We present the $\Mstar-Z_{\star}$ relation for our four different models in Fig. \ref{Ms_Zs}.  The luminosity-weighted stellar metallicity is  $Z_{\star}$ and is given in terms of mass fraction relative to the solar metallicity ($Z_{\odot} = 0.02$). Our model predictions are compared to the observational relation measured by \cite{Peng_2015}. This measure is based on more than 3900 nearby star-forming galaxy spectra extracted from the SDSS-DR7 survey.

Our predictions are also compared to the \cite{Gallazzi_2005} measurements. Based on the previous SDSS-DR2 survey, these investigators measured both stellar mass and light-weighted stellar metallicity for more than 53 000 nearby galaxies ($0.005<z<0.22$). This work is still a reference and is largely used for comparisons with models \citep[e.g.][]{Yates_2013, Lu_2015}.

We also compare our $\Mstar-Z_{\star}$ relations with a set of predictions from different SAMs \citep{Croton_2006,Somerville_2012,Lu_2011b}. The mean trend predicted by these SAMs was extracted from the recent paper of \cite{Lu_2014}. In these three SAMs the metal enrichment process is computed by assuming the instantaneous recycling approximation and fixed values for metal yields. 

\subsubsection{General behaviours}

The general behaviour our $\Mstar-Z_{\star}$ relations are in general agreement with those observed by \cite{Gallazzi_2005}. For stellar masses above $10^9\Msun$, the $\Mstar-Z_{\star}$ relations exhibit very similar trends for the four different models. In this range, the average $\Mstar-Z_{\star}$ relation, computed from both the whole sample and the SF galaxy sample, are indistinguishable. These strong similarities between models are mainly due to the computation of  a luminosity-weighted stellar metallicity. This quantity is mostly sensitive to young stellar populations: the youngest stellar populations, formed at $z\simeq 0.$, have  similar metallicity properties, even if they evolved previously in different conditions.

In our four models, the majority of our modelled galaxies in the range $10^9 <\Mstar<10^{10.5}\Msun$ are located on a plateau around $log(Z_{\star})\simeq -0.5$. This average value is consistent with those measured by \cite{Gallazzi_2005} around $10^9\Msun$. After this plateau, our models show a strong and prompt increase of the metallicity in the high-mass domain as measured by \cite{Gallazzi_2005}. After this strong increase in metallicity, the $\Mstar-Z_{\star}$ relation again becomes  constant with a level close to  solar metallicity. This average level is slightly smaller ($\simeq$ 0.1 dex) than that measured by \cite{Gallazzi_2005} in the SDSS data.

Unlike our models, the other SAMs do not reproduce the general trend observed in the SDSS-DR2 survey by \cite{Gallazzi_2005}. However, their average trend is similar to that measured by \cite{Peng_2015} in the SDSS-DR7 survey.

In our models, the increase starts above $10^{10.5}\Msun$. This transition mass is higher than that obtained by \citet{Gallazzi_2005} with a rapid increase in metallicity around $10^{9.5}\Msun$. In the mass range $10^{9.5} <\Mstar<10^{10.5}\Msun$, our simulated galaxies stay on the plateau and exhibit a metallicity (-0.3 dex) that is too low. In this specific range of mass, the rate of the metal enrichment is slower for our modelled galaxies than for the SDSS sample. The behaviour is similar in all  of our models, meaning that the causes of the discrepancy are not linked to the accretion scenario or to the star formation process. The causes of this delay might be related to the set of hypotheses used to model the link between the stars and  ISM. At any time, all the stellar populations previously formed re-inject enriched gas in the ISM. This ejected gas is instantaneously mixed with the pre-existing gas in the ISM. The gas ejected is therefore diluted and all the ISM has the same metallicity. In real galaxy discs, the stars are formed and evolve in regions of limited size. The gas ejected by the stellar population is not diluted with the total gas of the galaxy and remains  distributed around the star-forming region. Some stars are formed in regions with a higher metallicity than average. In our models, the metallicity signature of the gas ejected by the SN or AGN feedback process is similar to that of the ISM. We can presume that it is more difficult to eject metals, condensed in dust, than hydrogen. Even if this second point is essentially linked to the average gas metallicity, it could  also  impact  the stellar metallicity: a higher metallicity of the  ISM leads obviously to a higher stellar metallicity. 

\subsubsection{Impact of model prescriptions}

Even if the general trend of the $\Mstar-Z_{\star}$ relation is very similar in the four models, some differences can be observed. 

The impact of the non-star-forming gas recipe is visible in the low-mass domain ($\Mstar < 10^9\Msun$). In this range, the models $m_3$ and $m_4$ form a lower number of galaxies with $log(Z/\Zsun) > -0.5$. A decrease of the $\Mstar-Z_{\star}$ relation is observed for these models, while the average metallicity of low-mass objects is roughly constant in the two standard models ($m_1$ and $m_2$). With the non-star-forming gas reservoir, the star formation activity is strongly reduced and therefore the metal enrichment cycle is also strongly impacted.

The two different accretion scenarios lead to different $\Mstar-Z_{\star}$ relations below $10^{10}\Msun$. The $\Mstar-Z_{\star}$ relation predicted by models $m_1$ and $m_3$ (isotropic accretion) is above the prediction found for the models $m_2$ and $m_4$ (bimodal accretion). As expected the direct feeding with metal-free gas introduced  in $m_2$ and $m_4$ reduces the average stellar metallicity ($\simeq$ 0.1 dex). 

\subsection{Gas metallicity}
\label{gas_Z}

\begin{figure*}[t]
  \begin{center}
    \includegraphics[scale = 0.65]{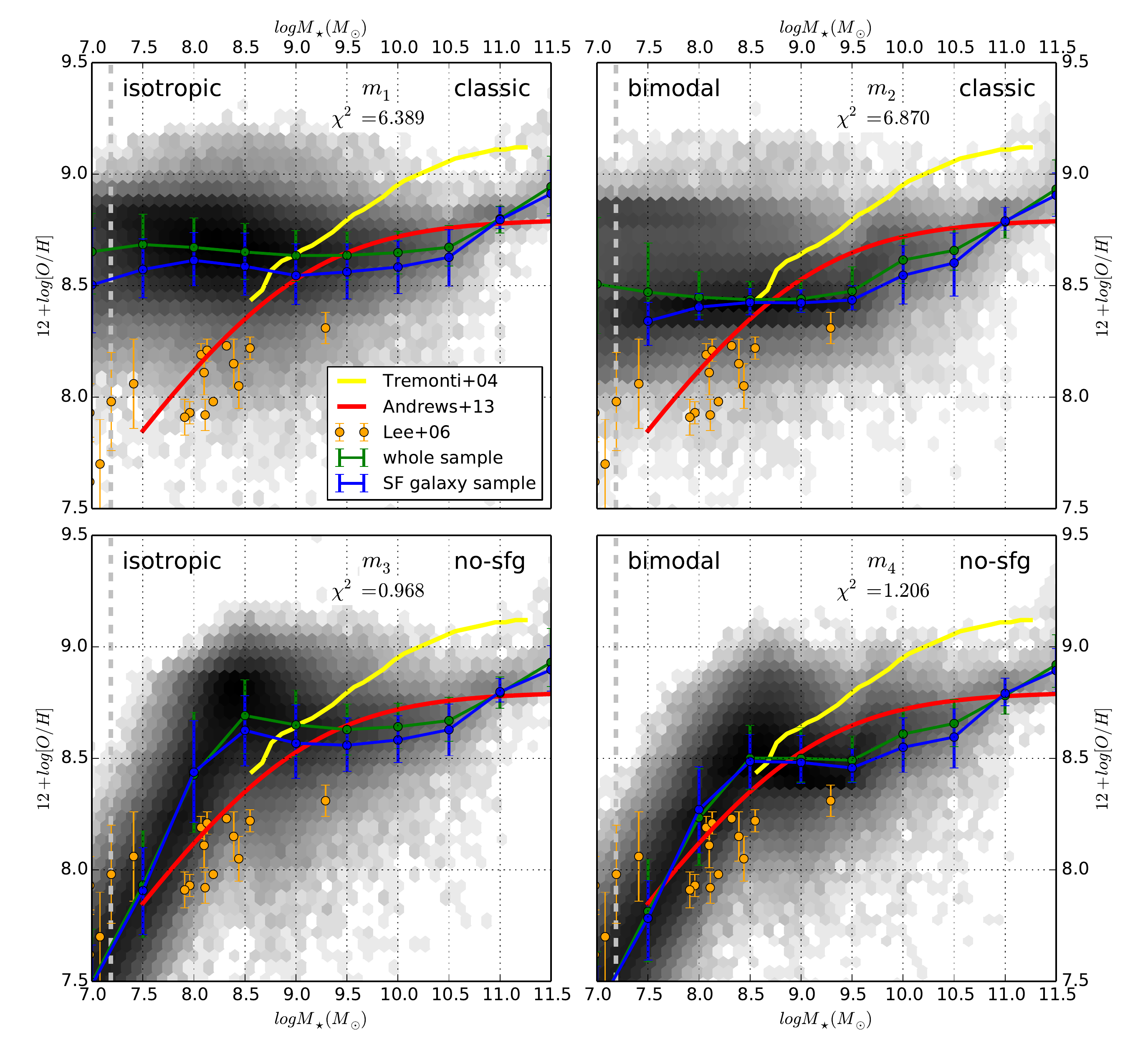}
        \caption{\tiny{$M_{\star}-Z_g$ relation. The gas metallicity $Z_g$ is given in terms of $12 + log_{10}\left(O/H\right)$; we assume $\Zsun = 8.94$ \citep{Karakas_2010}. The four distinct panels are distributed as in Fig. 5 (see Table \ref{model_definitions} for model references).  Same colours and grey scale are used for the distribution of modelled galaxies (all sample, median trend). Our predictions are compared to observational measurements from i) \cite{Tremonti_2004}: yellow solid line, ii) \cite{Andrews_2013}: solid red lines, and iii) \cite{Lee_2006}: orange points. The grey vertical dotted line represents the stellar mass resolution limit.}}
  \label{Ms_Zg}
  \end{center}
\end{figure*}

Oxygen is the most abundant element formed into stars. It can therefore be used as a proxy for all heavy elements and gas metallicity. We define the gas metallicity $Z_g$ as the number of oxygen atoms to hydrogen atoms with a logarithmic scale: $Z_g = 12 + log(O/H)$. In this formalism we adopt $\Zsun = 8.94$ \citep{Karakas_2010}.

Fig. \ref{Ms_Zg} shows the evolution of gas metallicity as a function of stellar mass. Our predictions are compared to various observational results: 
\begin{itemize}
        \item{\cite{Tremonti_2004} measured  both  stellar mass and gas-phase metallicity for more than 53.000 nearby ($z \simeq 0.1$) star-forming galaxies  from the  SDSS-DR2 survey \citep{York_2000}. Owing to our large  statistical set, this work
         has been the main observational set to be compared with models in more than ten years.}
        \item{\cite{Lee_2006} added infrared (IR) data coming from the $Spitzer$ Space Telescope to extend the $\Mstar-Z_g$ relations measured by \cite{Tremonti_2004} to the low stellar mass range (down to $\simeq 10^6\Msun$) }
        \item{ \cite{Andrews_2013}  measured the gas-phase metallicity of more than 200 000 nearby ($z \simeq 0.078$) star-forming galaxies from the SDSS-DR7 survey}
\end{itemize}

\cite{Tremonti_2004} and \cite{Andrews_2013}  used  different methods to measure metallicity:  i.e., the strong line and the direct method, respectively. The strong line method uses only the flux ratio of the strong lines. The metallicity of the gas is deduced  using empirical or theoretical calibrations. The direct method uses the flux ratio of auroral to strong lines to obtain a direct measure of the electron temperature of the gas. This temperature enables a better determination of the gas-phase metallicity. 

As discussed by \cite{Kewley_2008} or \cite{Moustakas_2010}, metallicities measured using theoretical calibrations based on strong lines are systematically larger than those obtained with the direct method. A gap between the metallicity measurements performed by \cite{Tremonti_2004} and \cite{Andrews_2013} is visible in Fig \ref{Ms_Zg} but found lower than the maximum systematic difference inferred by the different methods of measurements (up to $\simeq$ 0.7 dex; \cite{Kewley_2008}). Given all these uncertainties on metallicity measurements, we\ focus on the general trends instead of absolute values to constrain our models.

The SF galaxy sample and the whole sample exhibit very similar trends for all the models with,  on average, a slightly smaller metallicity for the SF galaxy sample  than for the whole sample.

While the $\Mstar-Z_{\star}$ relation has a very similar trend for all of the models, the $\Mstar-Z_g$ relation shows very different behaviour for the four different models. We observe a good agreement between the non-sfg scenario ($m_3$, $m_4$) and the \cite{Andrews_2013} measurements.

\subsubsection{The low-mass range: $\Mstar < 10^{9}\Msun$}

For the stellar masses below $\Mstar < 10^{9}\Msun$, the two different sets of star-forming models show distinct behaviours. The oxygen abundance is clearly over-predicted in the two classical models $m_1$ and $m_2$. The non-star-forming hypothesis used in the models $m_3$ and $m_4$ leads to a good agreement with the \cite{Lee_2006} and \cite{Andrews_2013} data points.

The oxygen excess found in $m_1$ and $m_2$ is directly linked to the low-mass galaxy excess observed in standard galaxy formation models and caused by  a  star formation activity that is too efficient. If the star formation process is not strongly regulated, the stellar mass assembly is too fast in low-mass structures and leads  to a systematic overproduction of stars. An excess of stars leads to an excess of metals. These results confirm the need for a strong regulation at the low-mass scale.

For a  given star formation process ([$m_1,m_2$] or [$m_3,m_4$]), the two  accretion scenarios lead to  similar trends. In the non-sfg scenario ($m_3,m_4$), the gas-phase metallicity increases faster   with an  isotropic accretion than in the bimodal accretion mode. In the low-mass domain ($\Mstar < 10^{8}\Msun$), the slope of the distribution is slightly larger for model $m_3$ than for model $m_4$. In addition, at $10^{8.5}\Msun$ the average gas-phase metallicity predicted by $m_3$ is larger than that  predicted by $m_4$ by 0.2 dex. The bimodal accretion mode allows  feeding of the galaxy with metal-free cosmological gas directly. The metal-enriched gas produced by the stars is diluted with the metal-free cosmological gas. In the isotropic accretion scenario, the newly accreted metal-free gas is mixed into the hot atmosphere with the metal-enriched gas coming from ejecta. The average gas-phase metallicity is therefore higher in the isotropic accretion scenario. With the  classical star formation models ($m_1,m_2$), the gas-phase metallicity is also slightly higher in the isotropic case.

\subsubsection{The intermediate-mass range: $10^{9} < \Mstar < 10^{10.5}\Msun$}

In this mass range, the predicted  gas-phase metallicities are very similar for the four models. The bimodal accretion mode leads again to a lower average metallicity for the gas phase as compared to the isotropic accretion mode. In all cases, the average metallicity of the models is consistent with that  measured by \cite{Andrews_2013}.

\subsubsection{The high-mass regime: $\Mstar > 10^{10.5}\Msun$}

In this range of mass, the four models give  also similar results. In all cases, oxygen abundances of our SF galaxies are consistent with \cite{Andrews_2013} metallicity measurements. However our relations still increase when a saturation of the metallicity  is observed by \cite{Andrews_2013} or \cite{Tremonti_2004}. This effect is likely to be due to our AGN feedback mechanism, which is not efficient enough (see Sect. 4.4 in \cite{Cousin_2015b} for more details).

\subsubsection{Our best model}

In the low-mass range ($\Mstar < 10^{9}\Msun$), standard star formation recipes lead to an overproduction of metals. These two models cannot be selected as our best model.  To identify this best model, we perform a reduced $\chi ^2$ test between both the \cite{Andrews_2013} and our SF galaxy sample mean trends. The reduced $\chi ^2$ values for each model are given in Fig \ref{Ms_Zg}. As mentioned previously, the two standard models obtain high $\chi^2$ values, they are therefore rejected. The model $m_3$, based on the non-sfg reservoir and the isotropic accretion process, is selected as the best model. Despite its better agreement in the low-mass regime, the $m_4$ model is rejected by the reduced $\chi ^2$ test\footnote{It should be possible to reduce this gap and obtain a better agreement with the model $m_4$ by tuning both cold stream and cooling efficiency parameters. Indeed in the bimodal accretion scenario, the transition between the two modes of accretion occurs in the intermediate range of masses ($10^{8.5} < \Mstar < 10^{10.5}$). The current set of parameters leads to a slight decrease of the net accretion rate onto the galaxy in this stellar mass range. A counterbalance of this decrease  increases the star formation activity and the average metallicity of the gas phase.}. The gap with \cite{Andrews_2013} measurements in the intermediate-mass range is larger than for model $m_3$. 
 
In the following sections, we use the model $m_3$ as our best model. 

\subsection{SFR, $\Mstar-Z_g$ relation}
\label{SFR-Zg-Ms}

In a seminal work, \cite{Mannucci_2010} (see also   \cite{Ellison_2008},  \cite{Lara-Lopez_2010})   introduced a fundamental relation between stellar mass, gas metallicity, and SFR in galaxies at low and high redshift as a generalisation of the mass metallicity relation. For a given stellar mass, metallicity globally decreases with increasing SFR. This anti-correlation between star formation activity and metallicity can be because of a modification of the gas content \citep[e.g.][]{Bothwell_2013,Lara-Lopez_2013}. The main actor could be low metallicity gas inflows which dilute metals in the gas phase, fuel star formation, and increase SFR \citep[e.g.][]{Zahid_2014}.
\begin{figure}[h]
  \includegraphics[scale=0.5]{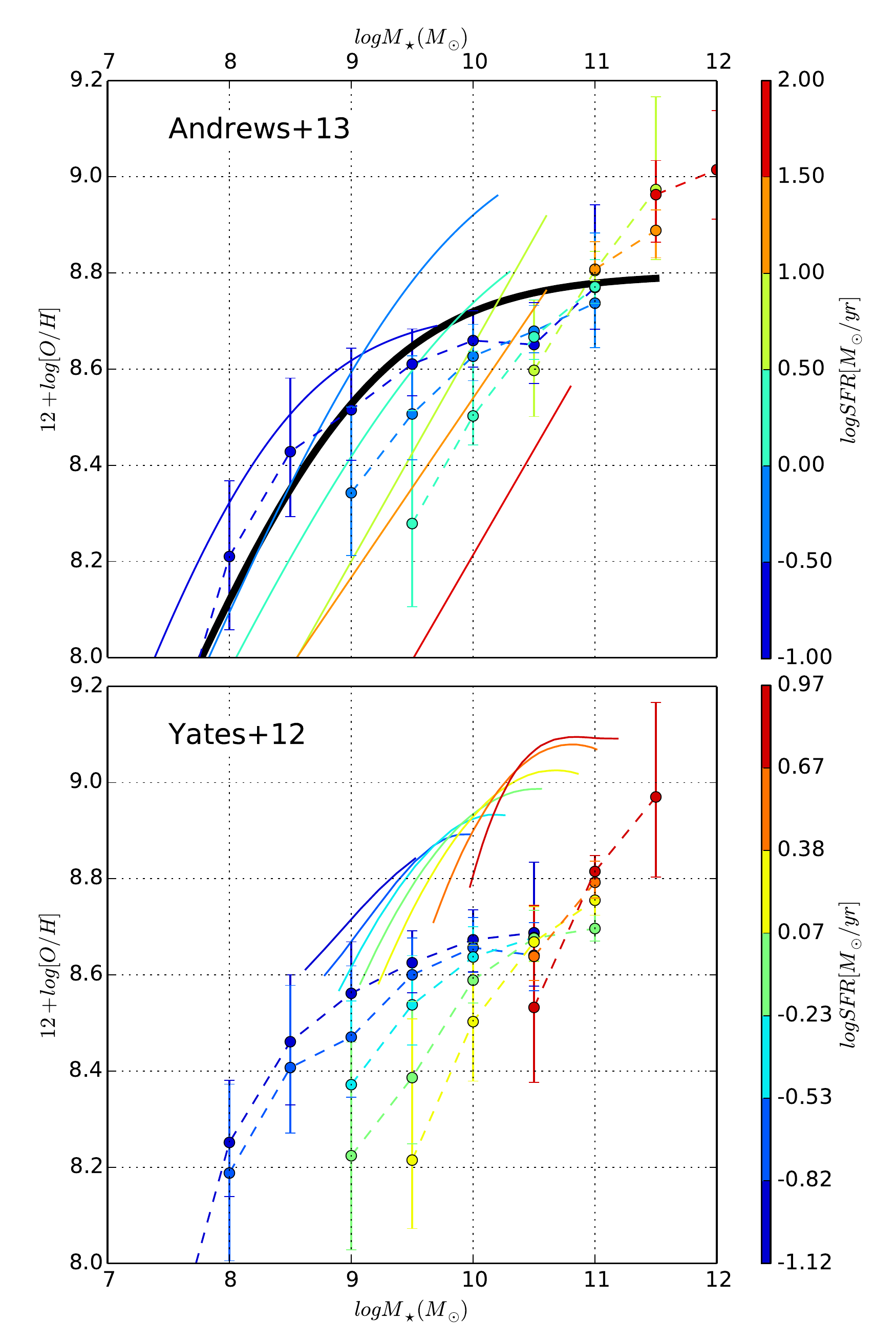}
  \caption{\tiny{SFR to $\Mstar-Z_g$ relation. Our best-model $m_3$ is compared to observational results from \cite{Andrews_2013} (upper panel) and \cite{Yates_2012} (bottom panel). In both cases, the original metallicity bins were respected. Solid lines show the observational results. The colour scale represents the SFR bin. The circles with error bars, connected by dashed lines, correspond to the median trend of our models. Error bars are computed using the first and third quartile of the distribution. In the upper panel, the solid black line represents the median trend of the $-\Mstar-Z_g$ relation.}}
  \label{Ms_Zg_SFR}
\end{figure}

Recently, \cite{Yates_2012} and \cite{Andrews_2013} revisited these relations with the SDSS-DR7 sample, their results are reported in Fig. \ref{Ms_Zg_SFR}. With a sample of more than 177 000 emission line galaxies \cite{Yates_2012} confirmed the dependence of metallicity on SFR at a fixed stellar mass, but with an opposite trend for low- and high-mass regimes with a transition around $\Mstar \simeq 10^{10}\Msun$. The relations of \cite{Andrews_2013} also shows a transition in their three first SFR-bins, which is similar to that reported by \cite{Yates_2012}. However, without measurements for stellar masses higher than $10^{10.5}\Msun$ in \cite{Andrews_2013}, it is difficult to give a definitive conclusion. \cite{Andrews_2013} use the direct method to compute gas-phase metallicity while in \cite{Yates_2012}, metallicities are calculated using a grid of photo-ionisation models \citep{Charlot_2001}. The difference in absolute levels of metallicity (Fig. \ref{Ms_Zg_SFR}) could be assigned to the different methods used. 

%  All these works find similar trends  but  the exact evolution observed as a function of the stellar mass is different. \cite{Mannucci_2010} observe an increase in gas metallicity with the stellar mass and an monotonic dependence of the gas metallicity with the SFR at fixed stellar mass. \cite{Yates_2012}, in their sample T2, do not extract an monotonic dependence betwwen $\Mstar-Z_g$ and SFR but they find a \textit{twist}. Their low-SFR galaxies, which are the most metal-rich at low masses, become the least metal-rich at high masses. Their sample T2 exhibits a SFR-dependence but with a transition between two distinct regimes around $\Mstar \simeq 10^{10}\Msun$. 

From the whole galaxy sample generated with our best model $m_3$, we extract galaxies with $-1.5 <log(SFR) < 2.0$ $\Msun/yr$  and  $8 <log(\Mstar) < 11.5$ $\Msun$. These ranges of SFR and stellar masses are similar to those used in \cite{Yates_2012} and \cite{Andrews_2013}. Our sample is then divided in (SFR, $\Mstar$) bins corresponding the samples defined by \cite{Yates_2012} and \cite{Andrews_2013} for the SFRs and with a width of 0.5 dex in $\Mstar$, respectively. For each of these $\Mstar$-SFR bins, the median metallicity of the gas is computed. In Fig. \ref{Ms_Zg_SFR}, our $\Mstar-Z_g$ relations are compared to the measurements of  \cite{Yates_2012} and \cite{Andrews_2013}

The $\Mstar-Z_g$ relation extracted from our best model $m_3$ shows a clear variation with SFR at fixed stellar mass. As in \ref{gas_Z} the absolute level measured by \cite{Andrews_2013} is in good agreement with our best model $m_3$. The dependence with the SFR is opposite for low- and high-mass regimes, as observed by \cite{Yates_2012}. The transition  occurs around $3\times 10^{10}\Msun$ for our model,  which is  slightly higher than  for  the \cite{Yates_2012} sample ($10^{10}\Msun$). This shift of our model towards high stellar masses also appears when we compare our model with \cite{Andrews_2013} results. Our model shows a good agreement with their average relation in the first SFR bin corresponding to $log(SFR) \in [-1.0 : -0.5]$. However we also observe a systematic shift for the bins of higher SFR ($log(SFR) \in [-0.5 : 1.0]$) of the order of 0.5 dex.  For $log(SFR) > 1.0$  the shift between our best model and the observational relation is smaller. 

Galaxies extracted from our best model appear too massive as compared to observations. This is consistent with the discrepancy found between the observed main sequence of galaxies and that extracted from our best model (see Fig. \ref{Main_sequence_z0}). As in the $\Mstar-Z_g$-SFR relation, the discrepancy for the main sequence is smaller at higher stellar mass ($\Mstar > 10^{10}$).

In summary, our SFR-$\Mstar-Z_g$ relation indicates that in a given SFR bin our galaxies are too massive as compared to observations but, it shows a clear variation with SFR at fixed stellar mass. As observed by \cite{Yates_2012}, the low-SFR galaxies that are the most metal rich at low masses become the least metal rich at high masses.

\begin{figure*}[t!]
  \begin{center}
    \includegraphics[scale = 0.53]{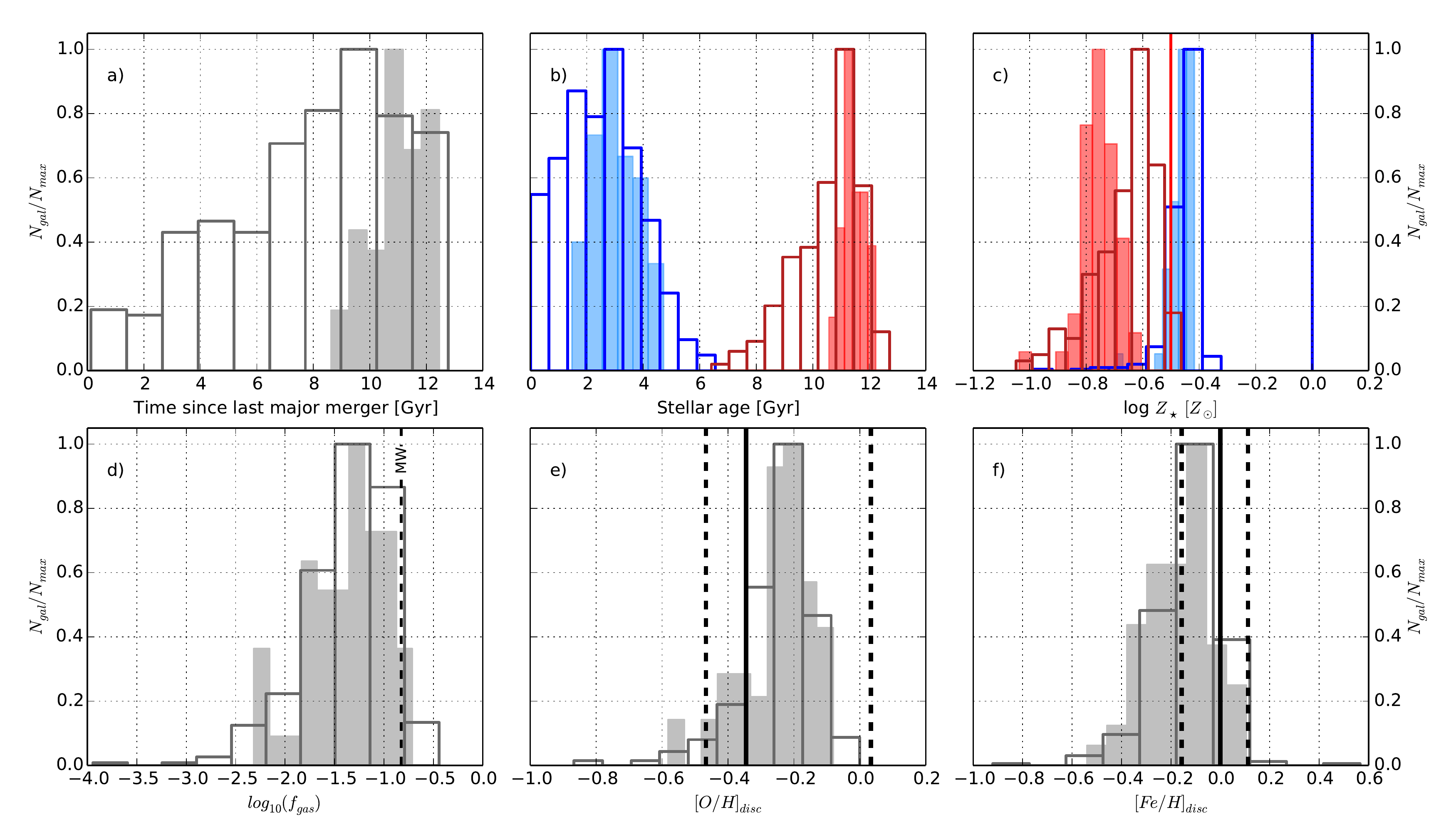}
        \caption{\tiny{Global properties for our Milky Way-like galaxy samples. Each histogram is normalised to the maximum number of objects contained in a bin. In each panel, the filled and  empty histograms are dedicated to our \textit{MW-sisters} sample and \textit{MW-cousins} sample, respectively. Panel \textbf{a)}: time since the last major merger. Panel \textbf{b)}: luminosity-weighted age of the disk (blue) and the bulge (red) stellar populations. Panel \textbf{c)} luminosity-weighted stellar metallicity of the disk (blue histogram) and bulge (red histogram). Our results are compared to \cite{Bensby_2014} (blue vertical line) for the stellar population of the disc and to \cite{Bensby_2013} (red vertical line) for the stellar population of the bulge. Panel \textbf{d)}: gas fraction; the dark vertical line indicates the gas fraction measured in the Milky Way by \cite{Yin_2009}. Panel \textbf{e)}: oxygen abundance $[O/H]_{disk}$ in the gas phase of the disk. Our predictions are compared to a compilation of oxygen abundances measured in recently formed open clusters \citep{Balser_2011, Maderak_2013}. Panel \textbf{f)}: iron abundance $[Fe/H]_{disk}$ in the gas-phase of the disk. Our predictions are compared to a compilation of iron abundances measured in recently formed open clusters \citep{Pancino_2010, Maderak_2013}. In panels \textbf{e} and \textbf{f}, the solid vertical black line and the two dashed vertical black lines represent the median, first, and third quartile of the data compilations, respectively.}}
  \label{MW_stats}
  \end{center}
\end{figure*}

\section{Focus on Miky Way-like galaxies}
\label{Miky_Way_like_galaxies}

In this section we study a sample of Milky Way-like galaxies generated with our best model $m_3$. The Milky Way is a SF galaxy. As a result of the number of constraints available, i.e.  stellar mass distribution, ages, metallicities, abundance gradient, a large number of works are dedicated to the evolution of the MW in term of stellar mass assembly and metallicity enrichment process. In this context, MW-like galaxies are good references to test our new model. In this section, we base our study on two groups of galaxies: the MW cousins and the MW sisters. The first group is selected using selection criteria based on the dark matter halo and total stellar masses. Then, by applying additional criteria we select the MW sisters as a sub-sample of the MW cousins. 

\subsection{\textit{Milky Way cousins}}

The measurements compiled by \cite{Yin_2009} or recently performed by \cite{Licquia_2015} indicate that the Milky Way has a total stellar mass between $3\times 10^{10}\Msun$ and $7\times 10^{10}\Msun$. In this stellar mass range, the agreement between the predicted $\Mstar$-SFR relation and the observed relation is satisfying (Fig. \ref{Main_sequence_z0}). Abundance matching methods \citep[e.g.][]{Behroozi_2010} allow us to define the following range: $7\times 10^{11}\Msun < M_h < 3\times 10^{12}\Msun$ for the dark matter halo hosting this range of stellar masses.

We randomly select 1440 halos in our simulation at z = 0 satisfying this dark matter halo mass criterion. For each one, the main branch of the merger tree is followed and saved with a high temporal resolution\footnote{For the main branch, the properties associated with the different components of the galaxy and of the dark matter halo are saved following the adaptive time-step algorithm used in eGalICS.}. The whole merger tree is also saved\footnote{The dark matter and baryonic properties of the main branch and all progenitor branches are saved at all the primary time-steps corresponding to the different snapshots of the initial N-body dark matter simulation.}  to follow dark matter and baryonic properties of all progenitors. 

From our initial selection of 1440 halos, we extract the \textit{MW-cousins} sample using the stellar mass criterion: $3\times 10^{10} < M_{\star} < 7\times 10^{10}\Msun$. This selection reduces our initial sample from 1440 to 336 objects. The majority of the rejected galaxies have a stellar mass between $1\times 10^{10}\Msun$ and $3\times 10^{10}\Msun$. In the range of dark matter halo mass selected, our modelled galaxies are therefore on average less massive than predicted by the abundance matching method.

\subsection{\textit{Milky Way sisters}}
 
In addition to its total stellar mass and dark matter halo mass, the Milky Way can be characterised by some other properties such as its SFR or its morphology. Based as previously on observational measurements we apply the following rules \citep{Yin_2009, Licquia_2015}:
\begin{itemize}
        \item{star formation activity: $1 < SFR < 5~\Msun/yr $ and}
        \item{morphology, defined as the ratio of the stellar mass hosted in disk to the total stellar mass of the galaxy: $0.6 < \frac{M_{\star,d}}{M_{\star}} < 0.8$.}
\end{itemize}

With these additional criteria, our \textit{MW-sisters} sample contains 56 objects. The majority of the rejected galaxies have a smaller disk to total mass ratio $\simeq 0.5$ and a SFR lower than  $1\Msun/yr$. 

\subsection{Global properties of Milky Way-like galaxies}

To analyse our selection, we focus on six quantities: the time elapsed since the last major merger,  stellar metallicity,  stellar age,   gas fraction, and the oxygen to hydrogen and iron to hydrogen relative abundances in the gas phase. The distributions of these quantities are shown in Fig. \ref{MW_stats} for both MW cousins and MW sisters. 

The panel a) of the Fig \ref{MW_stats} shows the distribution of the time elapsed since the last major merger. While for the \textit{MW-cousins} sample the width of the distribution is large, from 0.5 Gyr to 12.5 Gyrs, the distribution associated with the MW sisters is tightened around 10.5 Gyrs with a small scatter of 2 Gyrs. The morphology criterion that we used to select MW sisters induces a very high portion of the total stellar mass to be located in the disk of the galaxy and a small central bulge. Bulges are formed during major mergers; disk-dominated galaxies with a small bulge, such as MW-sister galaxies, encountered a major merger event  a long time ago. This kind of morphology is rare. The majority of our \textit{MW-cousins} galaxies are rejected from the \textit{MW-sisters} sample because they have a spheroidal morphology or a smaller disk than observed in our Milky Way. This old major merger leads to an old stellar population in the central bulge. This is confirmed by the distribution of the luminosity-weighted age and the metallicity of the stellar populations shown in panels b) and c) of Fig. 8, respectively. Stars in the central bulge of \textit{MW-sisters} galaxies have a luminosity-weighted ages between 10 and 12 Gyr. Their luminosity-weighted metallicity is also lower than that found for bulge stars of the \textit{MW-cousins} sample, which is consistent with an earlier formation for the $MW-sisters$ bulges. Stellar metallicities and ages of the stars located in the disk are similar in the two samples. The distributions associated with the MW cousins have a larger scatter than those predicted for \textit{MW-sisters} sample, while their mean values are very close. For both \textit{MW-cousins} and \textit{MW-sisters} samples, the stellar age and metallicity in the disk are representative of the recent story, and the past evolution has only a very small impact on them. We compare our distributions of stellar metallicities with those of \cite{Bensby_2014} and \cite{Bensby_2013} for the disk and bulge stellar populations, respectively. In comparison to these reference values our stellar metallicities are too low by 0.2 dex in the bulge and by 0.5 dex in the disk. The discrepency in terms of stellar metallicity in the bulge component is small given the uncertainty about yields (up to a factor 2). The low metallicity of our stellar disk, and therefore the gap between the measure of \cite{Bensby_2013} and our prediction, was already visible in Fig \ref{Ms_Zs}. Indeed, in Fig. \ref{Ms_Zs} our model predictions are offset by 0.5 dex as compared to the measurements of \citep{Gallazzi_2005}.

In the three lower panels of Fig. \ref{MW_stats}, the distributions of the gas fraction are plotted together with those of the oxygen and iron abundance relative to the hydrogen in the gas phase of the disk. The distributions associated with the \textit{MW-cousins} and  \textit{MW-sisters} sample exhibit similar average values and scatters. As for the stellar component of the discs, the past history does not have a strong impact on the gas properties.\\ 

We also compare our predictions to observed quantities. The measure of the gas fraction performed by \cite{Yin_2009} in the Milky-Way is reported in panel d). Our modelled galaxies have a smaller gas fraction than observed. The instantaneous SFR of our \textit{MW-sisters} galaxies is one of the criteria used to define the sample. The gas is therefore consumed with a rate similar to our Galaxy. A predicted gas fraction that is lower than observed seems therefore to indicate a lack of fresh gas. Our \textit{MW-sisters} galaxies are hosted by dark matter halo of $\simeq 10^{12}\Msun$ in which accretion is dominated by the cooling process of the hot gas phase. Our cooling efficiency has to be increased to reproduced the observed gas fraction.

In panel e) and panel f), we compare  our predicted $[O/H]_{disk}$ and $[Fe/H]_{disk}$ abundances, respectively, with a compilation of measures performed in recently formed open clusters ($<$3 Gyrs) \citep{Pancino_2010, Balser_2011, Maderak_2013}. The stars in these open clusters are young. Their chemical composition are therefore close to that of the gas in which they formed. 

Our predicted $[O/H]_{disk}$ abundance shows an average value that is slightly higher than observed. In contrast, the average value of our $[Fe/H]_{disk}$ abundance distribution is slightly lower than the observed value. The scatter of our predicted distribution is also smaller than observed. However, in the two cases, considering the dispersion and errors in measurements of the formation time of these open clusters, the predicted and observed abundances are in reasonable agreement.

In summary, MW sisters represents only 16\% of the \textit{MW-cousins} sample. Even if they host the same total stellar mass, the disk dominated morphology criterion strongly reduces the sample.
 
\subsection{Comparison with detailed chemical evolution models}
\label{MW-histories}

\begin{figure}[t]
 \includegraphics[scale =0.52]{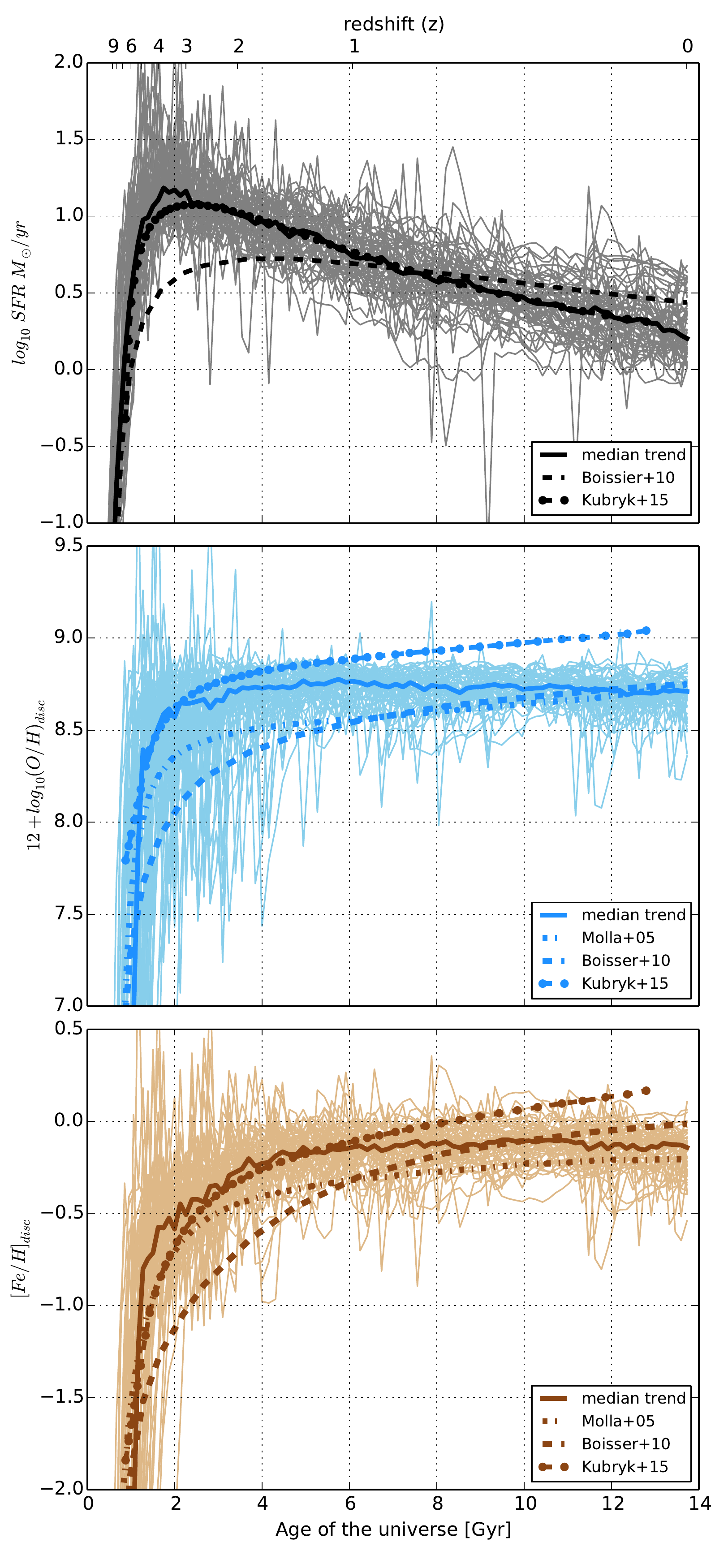}
 \caption{\tiny{Star formation and metal enrichment histories for \textit{MW-sisters} galaxies. The upper panel is dedicated to the star formation history, the central and  lower panels to the evolution of the oxygen abundance ($12+log(O/H)$) and of the iron abundance ($[Fe/H]$) in the disk. In each panel, the light colour curves show the individual evolution of our 54 \textit{MW-sisters} galaxies. The median trends of these evolutions are represented with a solid line. Our results are compared to the results of different detailed chemical evolution models of the MW galaxy: \cite{Molla_2005} (dot-dashed line), \cite{Boissier_2010} (dashed lined) and \cite{Kubryk_2015} (circles connected with a dashed line).}}
 \label{MW_model_timelines}
\end{figure}

In this section we present both the star formation and the metal enrichment histories of our 56 \textit{MW-sisters} galaxies. Based on their merger trees, we extract the median trends of both SFR and metal enrichment histories. The SFR history of each galaxy is computed by including the contribution of all its progenitors. Oxygen and iron abundances are followed only in the main disk's progenitor of the galaxies. 

Our median trends are compared to results coming from different detailed chemical evolution models aimed at reproducing MW properties in Fig. \ref{MW_model_timelines}
\begin{itemize}
        \item{\cite{Molla_2005} assume a universal rotation curve hypothesis and a multiphase chemical model and predict the radial mass distribution of 44 theoretical galaxies. They also provide the evolutions with time of oxygen and iron abundances. We compare our result with their model referenced as V=200 and N=4. This model assumes intermediate efficiencies for both: i) the conversion of the diffuse gas into molecular gas cloud and ii) the SFR. }
        \item{\cite{Boissier_2010} give the SFR and  chemical evolution of a set of disk galaxies. The grid of models is function of both the spin and  maximal rotational velocity of the dark matter host halo. The properties of these modelled disks are consistent with observed properties for redshift included in $0<z<1$.}
        \item{\cite{Kubryk_2015} base their models on a multiphase chemical evolution. They also include a description of radial migration based on the analysis of an N-body plus hydrodynamic simulation. This radial flow mixes gas of metal-poor regions into metal-rich regions. Their model reproduces current values of main global observables of the MW disk and bulge.}
\end{itemize}

This kind of models follows a classical approach for the mass assembly of the MW. These models do not follow the hierarchical formation of structures (as we do) but are constrained by additional observations (e.g. radial profile).

\subsubsection{Star formation rate history}

The median trend and individual SFR histories of our 56 \textit{MW-sisters} are shown in the upper panel of Fig. \ref{MW_model_timelines}. We observe strong short-time variations around the median trend. These variations are due to merger events or strong instantaneous baryonic accretion events that are not taken into account in \cite{Molla_2005}, \cite{Boissier_2010}, or \cite{Kubryk_2015}.

Beyond these variations, our median SFR history reaches a maximum after roughly 1 Gyr of evolution ($3.<z<4.$). Then the SFR slowly decreases until $z = 0$. 

This median SFR history is in agreement with the results of chemical evolution models. This history is especially consistent with that of \cite{Kubryk_2015}, which takes  radial transport inside the disk into account. Their predictions are compared favourably with a large suite of constraints in the Milky Way, such as age-metallicity relations. 

In the first 5 Gyrs of evolution, \cite{Boissier_2010} predict a lower SFR than both \cite{Kubryk_2015} and our model. In contrast, in the last 4 Gyrs of evolution,  \cite{Boissier_2010} predict a slightly larger SFR than our model even if their values are still in the dispersion generated by all our individual SFH histories. 

\subsubsection{Metal enrichment history}

In our model, the strong increase of the SFR, predicted just after the formation of the \textit{MW-sisters} galaxies, leads to a large production of metals. After only $\simeq$ 3 Gyr of evolution, both our median oxygen and iron abundances reach an asymptotic level. This saturation is not predicted by the three theoretical models, which instead predict a monotonic increase of both the oxygen and iron abundances. These different trends can be linked to the different feedback prescriptions and ejection processes assumed. In \cite{Boissier_2010} and \cite{Kubryk_2015} the new metals stay in the galaxy. These new metals are diluted with new fresh accreted gas, but their amount increases continuously. In our models the feedback processes produce winds that eject metals in the surrounding hot gas phase. Star formation and its associated feedback processes lead to an equilibrium between metals production and ejection. These two mechanisms explain the saturation in metal content and the asymptotic trend predicted by our models.

In the first Gyrs of evolution, the SFR predicted by \cite{Boissier_2010} is lower than the SFR predicted by \cite{Kubryk_2015} or by our models. Consequently the oxygen and  iron enrichment process is also slower during this period. However, the absolute value reached by \cite{Boissier_2010} at $z=0$ stays in good agreement with our results.

About \cite{Kubryk_2015} results, despite a SFR history similar to our, the oxygen and  iron abundances predicted reach larger values than in our \textit{MW-sisters} galaxies. 
As discussed previously, this result is consistent with a scenario with an accretion of fresh gas without gas ejection. 

Some other models have implemented a radial migration of gas and have tested the impact of these phenomenon on gas metallicity \citep[e.g.][]{Minchev_2014}. Their predictions for the evolution with time of $[Fe/H]$ are fully compatible with our results (\citealt{Minchev_2014} their Fig. 12).

The chemical evolution predicted by \cite{Molla_2005} shows the best agreements with our predictions for both oxygen and iron abundances. The behaviours (shapes and slopes) and the absolute levels at $z=0$ are especially consistent. 

In the next section, we extend the comparison of our predicted star formation histories of MW-like galaxies to observational measurements. 

\subsection{Star formation rate history of our MW-like galaxy samples.}

\begin{figure}[t]
 \includegraphics[scale =0.58]{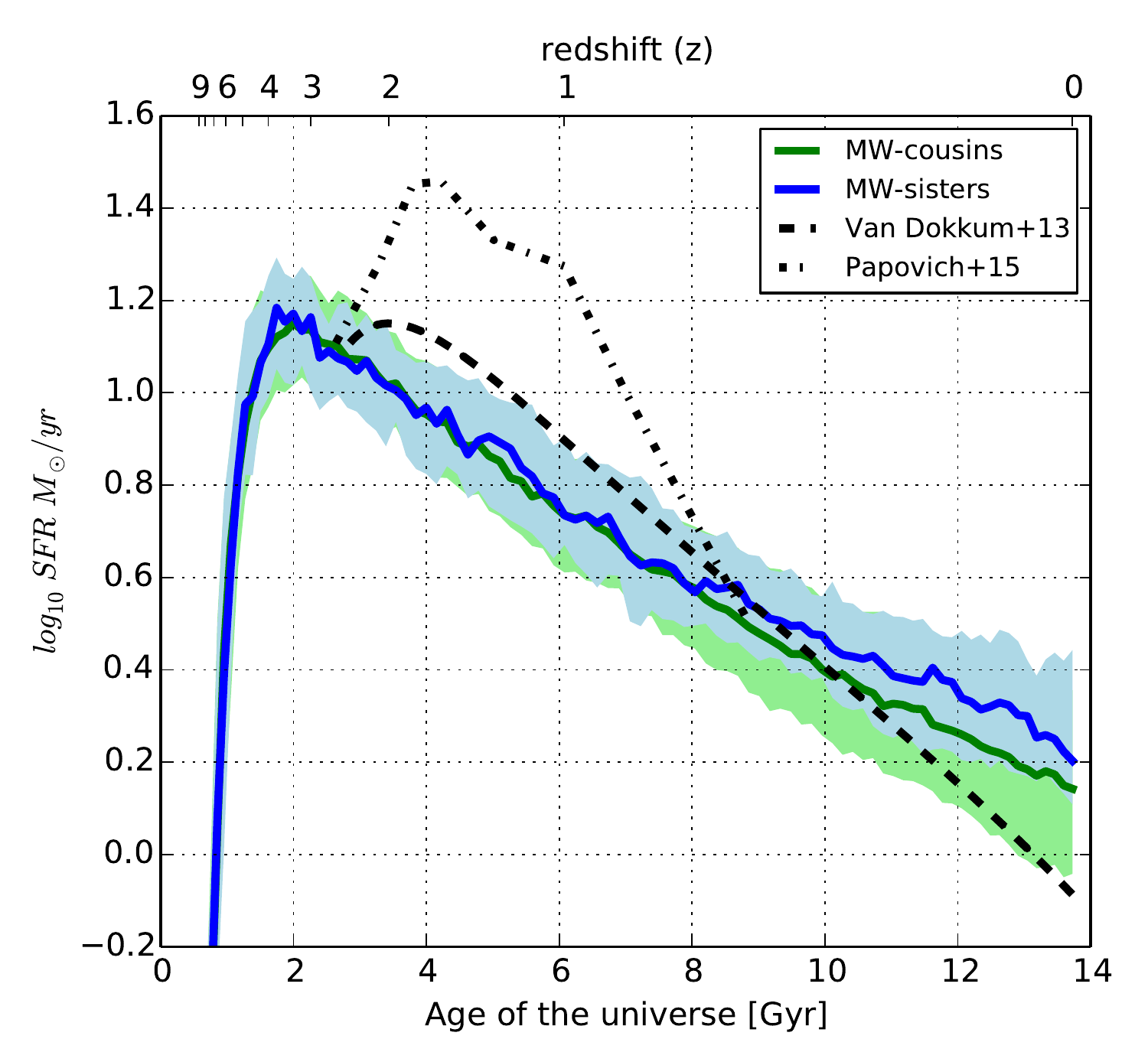}
 \caption{\tiny{Star formation histories. The green and blue solid lines show the median trends of the \textit{MW-cousins} sample and \textit{MW-sisters} sample, respectively. The associated coloured area is defined between the first and  third quartile of the distributions. Our results are compared with observational data of \cite{van_Dokkum_2013} (dashed line) and \cite{Papovich_2015} (dot dashed line).}}
 \label{MW_SFR_timelines}
\end{figure} 

\begin{figure*}[t!]
 \includegraphics[scale =0.65]{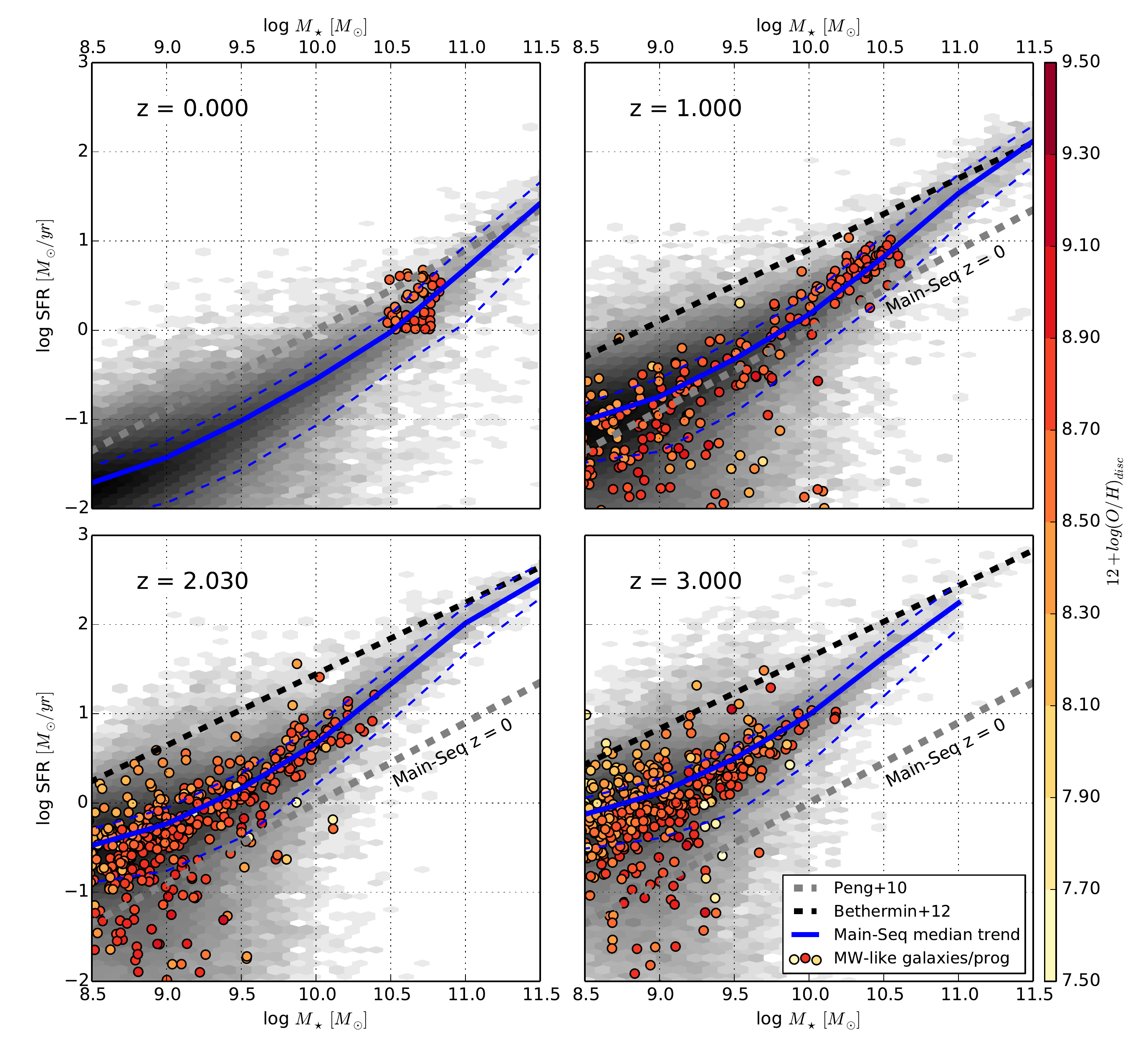}
 \caption{\tiny{Evolution with redshift of \textit{MW-sisters} progenitors onto the main sequence. The four panels are dedicated to four different redshift from $z=0.$ in the top left to $ z = 3.0$ in the bottom right. The grey scale shows the population of all star-forming (SFR $> 0$) galaxies extracted from our best model $m_3$, the blue solid line is the mean trend found for these galaxies, and the two blue dashed lines show the first and  third quartile of the distribution. Circles are \textit{MW-sisters} galaxies or their progenitors. The oxygen to hydrogen relative abundance in the disk's gas phase is colour\textit{}-coded. The grey dashed line represents the main sequence measured by \cite{Peng_2010} in the SDSS survey. The dark dashed line is the main sequence of \cite{Bethermin_2012c} for $z>=1$.}}
 \label{Main_sequence}
\end{figure*}

In this section, we present the median star formation histories of both \textit{MW-cousins} and \textit{MW-sisters} samples. Our results are compared in Fig \ref{MW_SFR_timelines} to the following observational measurements:
\begin{itemize}
        \item{\cite{van_Dokkum_2013} measure the stellar mass and the SFR of Milky Way-like galaxy progenitor out to $z \simeq 3$. Their analysis is based on 3D-HST and CANDELS Treasury surveys.}
        \item{\cite{Papovich_2015} use ZFOURGE, a deep medium-band, near-infrared imaging survey coupled with HST, Spitzer, and Herschel images, to measure the evolution in stellar mass, SFR, and also morphologies of Milky Way progenitors.}
\end{itemize}
In the two studies, the properties in stellar mass are deduced from abundance-matching techniques. As a prior, \cite{Papovich_2015} and \cite{van_Dokkum_2013} adopt a final stellar mass of $5\times 10^{10}\Msun$ for the Milky Way. We recall that our stellar mass range criterion is $3\times 10^{10} < \Mstar < 7\times 10^{10}$. 

In Fig \ref{MW_SFR_timelines}, the two median trends extracted from our two samples show very similar behaviours. We note a slight deviation in the four last Gyrs of evolution between the two samples. Our \textit{MW-sisters} galaxies are the most active; the SFR criterion generates a bias towards high SFR in the last four Gyrs of evolution. This result is consistent with the findings of \cite{Boissier_2010}. To match the stellar mass distribution of SF galaxy, these authors suggest that roughly 50\% of the Milky Way-like galaxies\footnote{In their study, the Milky Way-like galaxies are defined as galaxies with the same velocity as our Galaxy, i.e. 220km/s} have quenched their star formation since redshift 1 and therefore have a smaller SFR than the Milky Way. If this picture is correct, the Milky Way is more active than the majority of  galaxies with  similar stellar masses.

The star formation history measured by \cite{van_Dokkum_2013} is similar to what we obtain. Even if their maximum of activity is reached at lower redshift ($z \simeq 2.$), the absolute value of this peak is fully consistent with ours. Between $z = 4$ and $z = 0$, the slope of their star formation history is steeper than ours. Their measures are always included in our dispersion but reach the lowest values of our distribution at $z = 0$. 

The results of \cite{Papovich_2015} show a maximum for the star formation history around $1<z<2$, which is lower than the maximum that we find. This maximum is higher (0.3 dex) than both the \cite{van_Dokkum_2013} measurements and our predicted value. Between $z=2.$ and $z=0.5$, they measure a strong decrease of the SFR, which is steeper than the SFR we predict. The high SFR measured by  \cite{Papovich_2015} at $z\simeq 2$ is also in disagreement with \cite{van_Dokkum_2013} results, our model, and all the chemical evolution models presented previously. Indeed, this strong SFR peak would lead to a very strong metal enrichment process, which cannot be consistent with local ($z=0$) measurements. 

\subsection{Redshift evolution of the metallicity, SFR, and stellar mass of \textit{MW-sisters} progenitors}

\subsubsection{Metallicity of \textit{MW-sisters} progenitors}

The central panel of the Fig. \ref{MW_model_timelines} shows the evolution with time of the oxygen abundance predicted in the disk of the main progenitor of MW sisters. In this section, we extend this analysis at all the progenitors of our \textit{MW-sisters} galaxies.

Fig. \ref{Main_sequence} shows the relation $\Mstar-$SFR relation extracted from our best model $m_3$ at four different redhifts. In this figure, all individual points show a \textit{MW-sister} galaxy or a progenitor.The oxygen abundance in the disk's gas phase is colour-coded. For each redshift, we can see that for a very large percentage of the sample the oxygen abundance is high ($12 + log(O/H) \ge 8.5$). The metal enrichment process is therefore very efficient. We note that at $z\simeq 3$ the gas in the disks is already metal rich. By extending to all the progenitors, we confirm here the strong enrichment process observed in the main progenitor's disk in Fig. \ref{MW_model_timelines}. At $z\simeq 2$ and $z=3,$ we also confirm that on average at a given stellar mass, the higher the SFR, the lower the oxygen abundance. However, we can see some extreme points with low metallicities ($12 + log(O/H) \le 7.7$). These progenitors exhibit a low or very low star formation activity. These galaxies were extracted just after a merger event. During these events, the star formation activity is strongly increased (boost factor) and the gas is strongly consumed. After these violent events, in some cases, the star formation is strongly reduced because of a lack of fresh gas. These disks, without gas, accrete new gas, coming from the surrounding environment of the galaxy. This freshly accreted gas is generally poorer in metals in comparison to that previously consumed. After a merger, the metallicity can therefore strongly decrease. 

\subsubsection{Evolution onto the main sequence}

In this section we address the evolution on the main sequence of the progenitors of our MW-sister galaxies. The evolution of the $\Mstar-$SFR relation extracted from our best model $m_3$ is shown in Fig. \ref{Main_sequence}. In this figure, we also report the relation measured by \cite{Peng_2010} at $z\simeq 0.09$ and the evolution of the main sequence with redshift predicted by the empirical model of \cite{Bethermin_2012c} (for $z>=1$), which reproduces the observed galaxy main sequence well between $1.<z<3.$ \citep{Karim_2011,Rodighiero_2011,Whitaker_2012}. 

Measurements performed by \cite{Daddi_2007}, \cite{Karim_2011}, and \cite{Salmi_2012} indicate that, at fixed stellar mass, the average SFR increases with redshift. This trend is also predicted by our model. However, as in Fig. \ref{Main_sequence_z0}, we observe that, at all redshifts and in the low-mass regime, the observed SFR is lower than that obtained by \cite{Bethermin_2012c}. This is because of the strong regulation process applied on the star formation activity in low-mass structures. The non-sfg reservoir allows us to reproduce  the stellar mass function well\citep{Cousin_2015a} and gives satisfying results for the gas metallicity (Fig. \ref{gas_Z}), but at fixed stellar mass and in the low-mass regime the star formation activity is lower than observed. 

\subsubsection{Milky Way-sister galaxies and the main sequence}

At z = 0 a significant fraction of our MW sisters is above the mean trend of our predicted main sequence. As explained previously, SFRs of our \textit{MW-sisters} galaxies are higher than the majority of our simulated galaxy with a similar stellar mass (Fig. \ref{MW_model_timelines}). We also note that our MW sisters are close to the observed main sequence \citep{Peng_2010}, but are still mostly below the observed main sequence. 

At higher redshifts $z \ge 1$, the majority of the \textit{MW-sisters} progenitors are close to the $\Mstar-$SFR relation extracted from our whole sample. The progenitors of our \textit{MW-sisters} sample are therefore standard galaxies of our best model $m_3$. We note however that they are located below the observed main sequence at the corresponding redshift. 

These two previous points are consistent with the result of \cite{Heinis_2014}. Indeed they found that galaxies would have a stellar mass that is too large if they stay on the observed main sequence from high redshift to z = 0. Therefore the \textit{MW-sisters} progenitors cannot stay  their entire life on the observed main-sequence relation, but reach this relation only at redshift close to 0. Currently the Milky Way is on the observed local main sequence. This is based on the currently observed main sequence. If in reality the low-mass galaxies ($\Mstar < 10^{9}$) have lower SFR than currently measured \citep[e.g.][]{Whitaker_2014}, the observed main sequence should be closer to our predicted main sequence at high redshift. In these conditions the \textit{MW-sisters} progenitors could have evolved mainly onto the main sequence.

\section{Conclusions}
\label{conclusion}

In this work, we have focused our attention on the metallicity signature of the galaxies in the context of SAMs. To achieve this goal, we  added  a new chemodynamical extension to
our SAM, which follows the metal enrichment process into both the stellar population and gas phase. We  analysed the results of four different models based on two star formation scenarios and two accretion scenarios. 

We show that the four models lead to similar $\Mstar-Z_{\star}$ relations at $z\simeq 0$. The different scenarios of gas accretion and star formation have therefore no significant impact on this relationship. The shape of our average $\Mstar-Z_{\star}$ relations is in good agreement with observations, but we observe a systematic shift to the high masses. This shift should be linked to the set of simple rules used in the gas-stars coupling. 

In the $\Mstar-Z_g$ relationship, the two different models of star formation regulation lead to very different behaviours. In the low-mass regime, the oxygen abundance predicted by our classical model of star formation is too high compared with the measurements carried out in dwarf galaxies. Conversely, our model based on a strong regulation of the star formation shows better agreements. For a given star formation prescription, we note a slightly lower metallicity in the model based on bimodal accretion (i.e, metal-free cold-streams).

We then analysed the dependence of the $\Mstar-Z_{g}$ relation with the star formation activity. The absolute levels of our $\Mstar-Z_{g}-$SFR relations, extracted from our best model $m_3$, are in good agreement with the recent measurements of \cite{Andrews_2013}. In addition as observed by \cite{Yates_2012} our low-SFR modelled galaxies, which are the most metal-rich at low masses, become the least metal-rich at high masses. 

Based on dark matter halo mass and stellar mass, we  selected a group of MW cousins. From this first sample, we extracted the MW sisters by applying both SFR and morphological criteria. This last criterion strongly reduce the sample. The\textit{ MW-sisters \textup{sample}} represents only 16\% of the initial \textit{MW-cousins} sample. The SFR criterion also has an impact, we note that MW sisters are more active than their cousins. The median SFR history of our \textit{MW-sisters} sample is consistent with observational measurements and also shows very good agreement with detailed chemical evolution models. The high values of oxygen and iron abundances predicted by our best model at high redshifts $z\le 3.0$ shows that the metal enrichment process is very efficient.

We finally addressed the evolution onto the main sequence of the progenitors of our MW-sister galaxies. At fixed stellar mass, our model reproduces the observed increase with the redshift of the SFR. However our predicted main sequence is lower (by 0.5 dex) than the observed main sequence in the low-mass regime. Our \textit{MW-sisters} progenitors appear at high redshift, as the standard galaxies of our best model. However we have shown that our \textit{MW-sisters} galaxies start their evolution below the observed main sequence and progressively reach this observed relation at $z\simeq0$. 
 
\begin{acknowledgements}
Authors thank the Centre National d'Etude Spatial for its financial support. We thank Laura Portinari for very useful discussions. We thank the referee for his helpful report and his comments.
\end{acknowledgements}

\bibliographystyle{aa} 
\bibliography{aa5}

\newpage

\begin{appendix}

\section{From SSPs to realistic star formation histories: The history tables.}
\label{Histtables}

\subsection{The SSP ejecta rate table: $\EjTAB$}

Based on Eq. \ref{ejecta_rate_eq} we build a 3D (age, metallicity, and element) table: $\EjTAB$. We use 130 different age bins and 7 initial stellar metallicity bins. The chosen stellar ages values correspond to a sub-sample ($\Delta t_{min} = 1$ Myr) of the SSP-age list available in the BC03 stellar spectrum library. We assume that ejecta rates are constant between two successive stellar ages. The $\EjTAB$ is computed by assuming a given IMF. We use \cite{Chabrier_2003} but this table is available for \cite{Salpeter_1955}, \cite{Scalo_1998}, and \cite{Kroupa_2001} IMFs. In order to be consistent with the mass conservation, in each age bin, the instantaneous ejecta rate is normalised to $m_{\star} + m_{remnent} = 1\Msun$. 

The $\EjTAB$ is the first elementary block used to compute the stellar evolution and the metal enrichment process associated with the complex star formation history that occurs in our simulated galaxies.

\subsection{The stellar mass history table: $\SMHTAB$}
\label{sfh_tab}

For each galaxy, the stellar mass assembly is followed  using a 2D (age and metallicity)\ table: $\SMHTAB$. At each time of the evolution process, the cell $\SMHTAB [a_i,Z_{\star}^j]$ gives the mass of stars ($m = m_{\star} + m_{remnent}$) with an age: $a_i < age < a_{i+1}$ and with an initial stellar metallicity $Z_{\star}^j$. A clock \small\textbf{clk}$[a_i,Z_{\star}^j]$ is associated with each cell of the $\SMHTAB$. This clock runs as long as the cell contains stellar mass. The $\SMHTAB$ is the second elementary block used to compute the stellar evolution and the metal enrichment process associated with the complex star formation history.

\subsubsection{Increasing stellar mass}

During a star formation episode, a total mass of gas $M_g$ with a metallicity $Z_g$ (metal mass fractions) is transformed into new stars. The mass of each main-ISM element contained in this mass of gas is also known. The mass $M_g$ has to be added to the first age bin of $\SMHTAB$ and distributed into the different available initial stellar metallicity bins. We apply a set of rules:
\begin{itemize}
        \item{i) No stars can be formed with an initial stellar metallicity higher than the input average gas metallicity.}
        \item{ii) To be consistent with the composition of the total stellar ejecta associated with a given initial stellar metallicity, no interpolation between stellar metallicity bins is performed.}
        \item{iii) For each initial stellar metallicity bin, the initial chemical composition is fixed and given by \cite{Karakas_2010}.}
\end{itemize}
To be consistent with the amount of each element contained in the star-forming gas, the distribution of the gas between the different metallicity bins is carried out by maximising the mass affected in the bin that is closest to (but smaller than) the average gas metallicity $Z_g$. By following the previous rules, the chemical composition of the stellar mass actually formed is different from the input star-forming gas. To respect the mass conservation for each element followed, a residual mass is returned to the ISM\footnote{In average, the total mass re-injected into the ISM is close to $0.1\%$ of the initial star-forming gas mass and is always smaller than 5\%.}. 

\subsubsection{Evolution of the stellar population}

The evolution of a given stellar population is then performed following a adaptive time-steps scheme applying time-steps $\Delta t$ smaller or equal to 1 Myr. For each bin $[a_i,Z_{\star}^j]$, the evolution is carried out as follows:

\begin{itemize}
        \item{i) For each element tracked by the library, the mass $m_{elt}$, ejected by the stellar population hosted in the bin is re-incorporated into the ISM. This mass is given by 
                \begin{equation}
                        m_{elt} = \SMHTAB [a_i,Z_{\star}^j]\EjTAB [a_i,Z_{\star}^j,elt]\Delta t.
                \end{equation}}
        \item{ii) In a second time-step, the mass contained in a given bin of $\SMHTAB [a_i,Z_{\star}^j]$ can be (or not) transferred to the next (older) age bin: $a_{i+1}$. Depending on the clock linked to this bin, two cases are possible: 
                \begin{itemize}
                        \item{i) If \small\textbf{clk}$[a_i,Z_{\star}^j]+\Delta t < a_{i+1}$, no mass transfer is performed and we just add the current time-step to the clock,
                        \begin{tiny}
                                \begin{equation}
                                        \small\textbf{clk}[a_i,Z_{\star}^j] = \small\textbf{clk}[a_i,Z_{\star}^j]+\Delta t.
                        \end{equation}
                        \end{tiny}}
                        \item{ii) If $\small\textbf{clk}[a_i,Z_{\star}^j]+\Delta t > a_{i+1}$, all the mass in the bin $[a_i,Z_{\star}^j]$ is transferred into the next age bin $[a_{i+1},Z_{\star}^j]$. This transfer of mass towards the next age bin leads to a change of its clock. The new value is set to
                        \begin{tiny}
                                \begin{equation}
                                        \begin{split}
                                                clk[a_{i+1},Z_{\star}^j] = & \frac{m[a_i,Z_{\star}^j]\times(clk[a_i,Z_{\star}^j]+\Delta t)}{m[a_i,Z_{\star}^j]+m[a_{i+1},Z_{\star}^j]}  \\
                                                                                        & + \frac{m[a_{i+1},Z_{\star}^j]\times clk[a_{i+1},Z_{\star}^j]}{m[a_i,Z_{\star}^j]+m[a_{i+1},Z_{\star}^j]}
                                        \end{split}
                                .\end{equation}
                        \end{tiny}}
        \end{itemize}}
\end{itemize}

The $\SMHTAB$ is updated starting with the oldest age bin containing mass and, afterwards, by browsing all younger bins. The evolution procedure ends by the younger age bin: 0 Myr in which a new star formation episode can be taken into account.

\subsection{Instantaneous SN rate: $\SNTAB$}
\label{inst_sn_rate}

We introduce a realistic SNII + SNIa even rate. This rate is used to compute SN feedback. 

\begin{figure}[h]
        \begin{center}
        \includegraphics[scale=0.55]{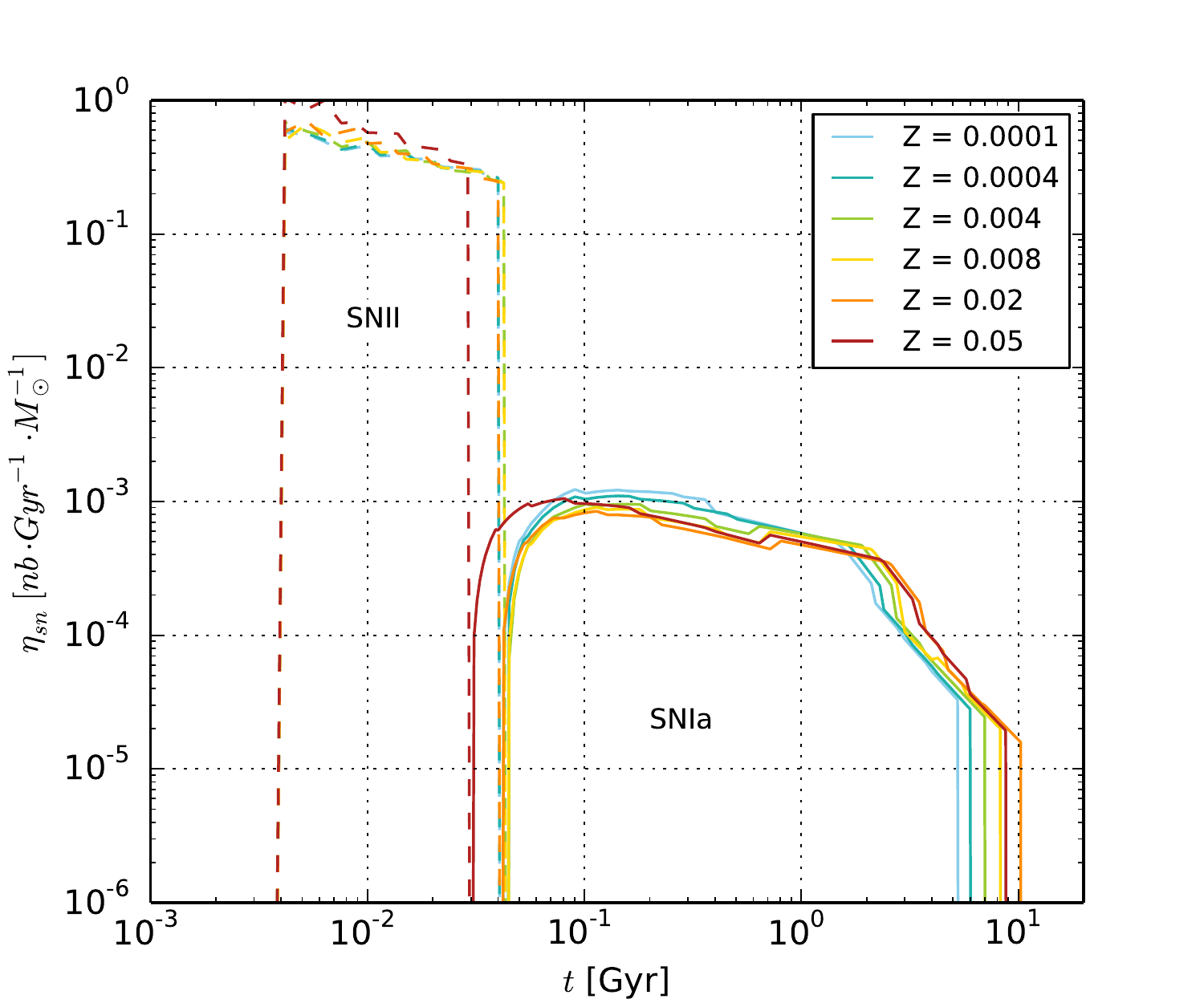}
        \caption{\tiny{SN event rates computed for a single stellar population of $1\Msun$ distributed through a Chabrier IMF. Type II supernovae rates are shown with dotted lines. Type Ia supernovae rates are plotted with solid lines. The different stellar metallicities are materialised by the different colours.}}
        \label{SN_rates}
        \end{center}
\end{figure}

Based on Eqs. \ref{eta_AGB_SNII} and \ref{eta_SNIa}, we build a 2D (age and metallicity) table: $\SNTAB$: 
\begin{equation}
                \SNTAB[a_i,Z_{\star}^j] = \left.\eta_{\star}(t=a_i,Z_{\star}^j)\right|_{m>9\Msun} + \eta_{snIa}(t=a_i,Z_{\star}^j)
.\end{equation}
In this definition the first term $\left.\eta_{\star}(t=a_i,Z_{\star}^j)\right|_{m>9\Msun}$ is the death rate of a massive star applied for a massive star with a lifetime $\tau_m = a_i$. The second term, $\eta_{snIa}(t=a_i,Z_{\star}^j),$ is the instantaneous SNIa rate computed at a time $t=a_i$. We assume that the global SN rate is constant between two successive ages. As for the instantaneous ejecta rate, at a given age, the SN rate is normalised to $m_{\star} + m_{remnent} = 1\Msun$.

Figure \ref{SN_rates} shows the instantaneous SNII and SNIa event rates for different stellar metalicities. This SN rates are associated with a SSP of $1\Msun$ distributed through a Chabrier IMF.  

By using $\SNTAB$ and $\SMHTAB$ the instantaneous SN rate, $\eta_{sn}$, used in Eq. \ref{sn_feedback_eq} is simply given by
\begin{equation}
                \eta_{sn} =  \sum_{i}\sum_{j}\SNTAB[a_i,Z_{\star}^j]\SMHTAB [a_i,Z_{\star}^j]
.\end{equation}

\section{Data repository}

\subsection{Model outputs}
Outputs from all models presented here are available on CDS platform \verb? http://cdsweb.u-strasbg.fr?. Data are distributed under \verb?fits?-formatted files and are therefore compatible with the \verb?TOPCAT? software (\verb?http://www.star.bris.ac.uk/~mbt/topcat/?).\\ 

We also provide the median star formation history, median gas accretion history, and metal enrichment histories (12 + log[O/H] and [Fe/H]) associated with our \textit{MW-sisters} sample (Fig. \ref{MW_SFR_timelines}). These data are saved in \verb?ascii?-formated files.

\subsection{Chemodynamical library}

The study presented here is mainly based on a new chemodynamical library. We provide a set of two files associated with this library:
\begin{itemize}
        \item{\verb?mass_loss_rates_IMF.fits? files, which contain the ejecta rate [$\Msun/Gyr$] associated with a SSP of $1\Msun$ for the six different main-ISM elements followed: $^1H$, $^4He$, $^{12}C$, $^{14}N$, $^{16}O,$ and $^{56}Fe$. The total ejected mass and metal ejected mass are also provided (Eq. \ref{ejecta_rate_eq} and Fig. \ref{Ejecta_rates_plot}).}
        \item{\verb?SN_rates_IMF.fits? files, which contain the total SN rate (SNII + SNIa [$nb/Gyr$]) associated with a SSP of $1\Msun$ and with the individual contribution of SNII and SNIa (Fig. \ref{SN_rates}).}
\end{itemize}
These files are available for four different IMFs: Salpeter+55, Chabrier+03, Kroupa+93 and Scalo+98. Both ejecta rates and SN rates are computed for the complete list of stellar ages provided in the BC03 spectra library. They are saved in \verb?fits?-formated files and structured with different extensions corresponding to the different initial stellar metallicity bins.

\end{appendix}

\end{document}